%Paper: hep-th/9112046
%From: Kanehisa Takasaki <TAKASAKI%JPNYITP.BITNET@pucc.Princeton.EDU>
%Date: Wed, 18 Dec 91 14:35:59 JST
%Date (revised): Mon, 01 Feb 93 12:24:32 JST

%%%%% SDiff(2) KP hierarchy, by Kanehisa Takasaki and Takashi Takebe
%%%%% Proceedings of RIMS Research Project 91 "Infinite Analysis"
%%%%% International Journal of Modern Physics, Suppl. 1B (1992)
%%%%% Errors in the published version are corrected
%%%%%
%%%%%%%%%%% macros extracted from "vanilla.sty" %%%%%%%%%%%%%%%
\catcode`\@=11
\font\tensmc=cmcsc10      %change to CM fonts 3-31-87
%\font\tensmc=amcsc10
\def\smc{\tensmc}

\def\eat@#1{}
\mathchardef\prime@="0230
\def\prime{{{}\prime@{}}}
\def\prim@s{\prime@\futurelet\next\pr@m@s}

\def\,{\relax\ifmmode\mskip\thinmuskip\else\thinspace\fi}
\def\!{\relax\ifmmode\mskip-\thinmuskip\else\negthinspace\fi}
\def\frac#1#2{{#1\over#2}}
\def\dfrac#1#2{{\displaystyle{#1\over#2}}}

\def\:{\nobreak\hskip.1111em{:}\hskip.3333em plus .0555em\relax}
\def\intic@{\mathchoice{\hskip5\p@}{\hskip4\p@}{\hskip4\p@}{\hskip4\p@}}
\def\negintic@
 {\mathchoice{\hskip-5\p@}{\hskip-4\p@}{\hskip-4\p@}{\hskip-4\p@}}
\def\intkern@{\mathchoice{\!\!\!}{\!\!}{\!\!}{\!\!}}
\def\intdots@{\mathchoice{\cdots}{{\cdotp}\mkern1.5mu
    {\cdotp}\mkern1.5mu{\cdotp}}{{\cdotp}\mkern1mu{\cdotp}\mkern1mu
      {\cdotp}}{{\cdotp}\mkern1mu{\cdotp}\mkern1mu{\cdotp}}}
\newcount\intno@
\def\iint{\intno@=\tw@\futurelet\next\ints@}
\def\iiint{\intno@=\thr@@\futurelet\next\ints@}
\def\iiiint{\intno@=4 \futurelet\next\ints@}
\def\idotsint{\intno@=\z@\futurelet\next\ints@}
\def\ints@{\findlimits@\ints@@}
\newif\iflimtoken@
\newif\iflimits@
\def\findlimits@{\limtoken@false\limits@false\ifx\next\limits
 \limtoken@true\limits@true
   \else\ifx\next\nolimits\limtoken@true\limits@false
    \fi\fi}
\def\multintlimits@{\intop\ifnum\intno@=\z@\intdots@
  \else\intkern@\fi
    \ifnum\intno@>\tw@\intop\intkern@\fi
     \ifnum\intno@>\thr@@\intop\intkern@\fi\intop}
\def\multint@{\int\ifnum\intno@=\z@\intdots@\else\intkern@\fi
   \ifnum\intno@>\tw@\int\intkern@\fi
    \ifnum\intno@>\thr@@\int\intkern@\fi\int}
\def\ints@@{\iflimtoken@\def\ints@@@{\iflimits@
   \negintic@\mathop{\intic@\multintlimits@}\limits\else
    \multint@\nolimits\fi\eat@}\else
     \def\ints@@@{\multint@\nolimits}\fi\ints@@@}
\def\Sb{_\bgroup\vspace@
        \baselineskip=\fontdimen10 \scriptfont\tw@
        \advance\baselineskip by \fontdimen12 \scriptfont\tw@
        \lineskip=\thr@@\fontdimen8 \scriptfont\thr@@
        \lineskiplimit=\thr@@\fontdimen8 \scriptfont\thr@@
        \Let@\vbox\bgroup\halign\bgroup \hfil$\scriptstyle
            {##}$\hfil\cr}
\def\endSb{\crcr\egroup\egroup\egroup}
\def\Sp{^\bgroup\vspace@
        \baselineskip=\fontdimen10 \scriptfont\tw@
        \advance\baselineskip by \fontdimen12 \scriptfont\tw@
        \lineskip=\thr@@\fontdimen8 \scriptfont\thr@@
        \lineskiplimit=\thr@@\fontdimen8 \scriptfont\thr@@
        \Let@\vbox\bgroup\halign\bgroup \hfil$\scriptstyle
            {##}$\hfil\cr}
\def\endSp{\crcr\egroup\egroup\egroup}
\def\Let@{\relax\iffalse{\fi\let\\=\cr\iffalse}\fi}
\def\vspace@{\def\vspace##1{\noalign{\vskip##1 }}}
\def\aligned{\,\vcenter\bgroup\vspace@\Let@\openup\jot\m@th\ialign
  \bgroup \strut\hfil$\displaystyle{##}$&$\displaystyle{{}##}$\hfil\crcr}
\def\endaligned{\crcr\egroup\egroup}
\def\matrix{\,\vcenter\bgroup\Let@\vspace@
    \normalbaselines
  \m@th\ialign\bgroup\hfil$##$\hfil&&\quad\hfil$##$\hfil\crcr
    \mathstrut\crcr\noalign{\kern-\baselineskip}}
\def\endmatrix{\crcr\mathstrut\crcr\noalign{\kern-\baselineskip}\egroup
                \egroup\,}
\newtoks\hashtoks@
\hashtoks@={#}
\def\format{\crcr\egroup\iffalse{\fi\ifnum`}=0 \fi\format@}
\def\format@#1\\{\def\preamble@{#1}%
  \def\c{\hfil$\the\hashtoks@$\hfil}%
  \def\r{\hfil$\the\hashtoks@$}%
  \def\l{$\the\hashtoks@$\hfil}%
  \setbox\z@=\hbox{\xdef\Preamble@{\preamble@}}\ifnum`{=0 \fi\iffalse}\fi
   \ialign\bgroup\span\Preamble@\crcr}
\def\pmatrix{\left(\matrix} \def\endpmatrix{\endmatrix\right)}

\def\cases{\left\{\,\vcenter\bgroup\vspace@
     \normalbaselines\openup\jot\m@th
       \Let@\ialign\bgroup$##$\hfil&\quad$##$\hfil\crcr
      \mathstrut\crcr\noalign{\kern-\baselineskip}}
\def\endcases{\endmatrix\right.}
\newif\iftagsleft@
\tagsleft@true
\def\TagsOnRight{\global\tagsleft@false}
\def\tag#1$${\iftagsleft@\leqno\else\eqno\fi
 \hbox{\def\pagebreak{\global\postdisplaypenalty-\@M}%
 \def\nopagebreak{\global\postdisplaypenalty\@M}\rm(#1\unskip)}%
  $$\postdisplaypenalty\z@\ignorespaces}
\interdisplaylinepenalty=\@M
\def\allowdisplaybreak@{\def\allowdisplaybreak{\noalign{\allowbreak}}}
\def\displaybreak@{\def\displaybreak{\noalign{\break}}}
\def\align#1\endalign{\def\tag{&}\vspace@\allowdisplaybreak@\displaybreak@
  \iftagsleft@\lalign@#1\endalign\else
   \ralign@#1\endalign\fi}
\def\ralign@#1\endalign{\displ@y\Let@\tabskip\centering
   \halign to\displaywidth
     {\hfil$\displaystyle{##}$\tabskip=\z@&$\displaystyle{{}##}$\hfil
       \tabskip=\centering&\llap{\hbox{(\rm##\unskip)}}\tabskip\z@\crcr
             #1\crcr}}
\def\lalign@
 #1\endalign{\displ@y\Let@\tabskip\centering\halign to \displaywidth
   {\hfil$\displaystyle{##}$\tabskip=\z@&$\displaystyle{{}##}$\hfil
   \tabskip=\centering&\kern-\displaywidth
        \rlap{\hbox{(\rm##\unskip)}}\tabskip=\displaywidth\crcr
               #1\crcr}}
\def\overrightarrow{\mathpalette\overrightarrow@}
\def\overrightarrow@#1#2{\vbox{\ialign{$##$\cr
    #1{-}\mkern-6mu\cleaders\hbox{$#1\mkern-2mu{-}\mkern-2mu$}\hfill
     \mkern-6mu{\to}\cr
     \noalign{\kern -1\p@\nointerlineskip}
     \hfil#1#2\hfil\cr}}}
\def\overleftarrow{\mathpalette\overleftarrow@}
\def\overleftarrow@#1#2{\vbox{\ialign{$##$\cr
     #1{\leftarrow}\mkern-6mu\cleaders
      \hbox{$#1\mkern-2mu{-}\mkern-2mu$}\hfill
      \mkern-6mu{-}\cr
     \noalign{\kern -1\p@\nointerlineskip}
     \hfil#1#2\hfil\cr}}}
\def\overleftrightarrow{\mathpalette\overleftrightarrow@}
\def\overleftrightarrow@#1#2{\vbox{\ialign{$##$\cr
     #1{\leftarrow}\mkern-6mu\cleaders
       \hbox{$#1\mkern-2mu{-}\mkern-2mu$}\hfill
       \mkern-6mu{\to}\cr
    \noalign{\kern -1\p@\nointerlineskip}
      \hfil#1#2\hfil\cr}}}
\def\underrightarrow{\mathpalette\underrightarrow@}
\def\underrightarrow@#1#2{\vtop{\ialign{$##$\cr
    \hfil#1#2\hfil\cr
     \noalign{\kern -1\p@\nointerlineskip}
    #1{-}\mkern-6mu\cleaders\hbox{$#1\mkern-2mu{-}\mkern-2mu$}\hfill
     \mkern-6mu{\to}\cr}}}
\def\underleftarrow{\mathpalette\underleftarrow@}
\def\underleftarrow@#1#2{\vtop{\ialign{$##$\cr
     \hfil#1#2\hfil\cr
     \noalign{\kern -1\p@\nointerlineskip}
     #1{\leftarrow}\mkern-6mu\cleaders
      \hbox{$#1\mkern-2mu{-}\mkern-2mu$}\hfill
      \mkern-6mu{-}\cr}}}
\def\underleftrightarrow{\mathpalette\underleftrightarrow@}
\def\underleftrightarrow@#1#2{\vtop{\ialign{$##$\cr
      \hfil#1#2\hfil\cr
    \noalign{\kern -1\p@\nointerlineskip}
     #1{\leftarrow}\mkern-6mu\cleaders
       \hbox{$#1\mkern-2mu{-}\mkern-2mu$}\hfill
       \mkern-6mu{\to}\cr}}}
\def\sqrt#1{\radical"270370 {#1}}
\def\dots{\relax\ifmmode\let\next=\ldots\else\let\next=\tdots@\fi\next}
\def\tdots@{\unskip\ \tdots@@}
\def\tdots@@{\futurelet\next\tdots@@@}
\def\tdots@@@{$\mathinner{\ldotp\ldotp\ldotp}\,
   \ifx\next,$\else
   \ifx\next.\,$\else
   \ifx\next;\,$\else
   \ifx\next:\,$\else
   \ifx\next?\,$\else
   \ifx\next!\,$\else
   $ \fi\fi\fi\fi\fi\fi}
\def\text{\relax\ifmmode\let\next=\text@\else\let\next=\text@@\fi\next}
\def\text@@#1{\hbox{#1}}
\def\text@#1{\mathchoice
 {\hbox{\everymath{\displaystyle}\def\textfonti{\the\textfont1 }%
    \def\textfontii{\the\textfont2 }\textdef@@ T#1}}
 {\hbox{\everymath{\textstyle}\def\textfonti{\the\textfont1 }%
    \def\textfontii{\the\textfont2 }\textdef@@ T#1}}
 {\hbox{\everymath{\scriptstyle}\def\textfonti{\the\scriptfont1 }%
   \def\textfontii{\the\scriptfont2 }\textdef@@ S\rm#1}}
 {\hbox{\everymath{\scriptscriptstyle}%
   \def\textfonti{\the\scriptscriptfont1 }%
   \def\textfontii{\the\scriptscriptfont2 }\textdef@@ s\rm#1}}}
\def\textdef@@#1{\textdef@#1\rm \textdef@#1\bf
   \textdef@#1\sl \textdef@#1\it}

\def\textdef@#1#2{%
 \def\next{\csname\expandafter\eat@\string#2fam\endcsname}%
\if S#1\edef#2{\the\scriptfont\next\relax}%
 \else\if s#1\edef#2{\the\scriptscriptfont\next\relax}%
 \else\edef#2{\the\textfont\next\relax}\fi\fi}
\scriptfont\itfam=\tenit \scriptscriptfont\itfam=\tenit
\scriptfont\slfam=\tensl \scriptscriptfont\slfam=\tensl
\mathcode`\0="0030
\mathcode`\1="0031
\mathcode`\2="0032
\mathcode`\3="0033
\mathcode`\4="0034
\mathcode`\5="0035
\mathcode`\6="0036
\mathcode`\7="0037
\mathcode`\8="0038
\mathcode`\9="0039
\def\Cal{\relax\ifmmode\let\next=\Cal@\else
    \def\next{\errmessage{Use \string\Cal\space only in %
      math mode}}\fi\next}
    \def\Cal@#1{{\fam2 #1}}
\def\bold{\relax\ifmmode\let\next=\bold@\else
    \def\next{\errmessage{Use \string\bold\space only in %
      math mode}}\fi\next}
    \def\bold@#1{{\fam\bffam #1}}
\mathchardef\Gamma="0000
\mathchardef\Delta="0001
\mathchardef\Theta="0002
\mathchardef\Lambda="0003
\mathchardef\Xi="0004
\mathchardef\Pi="0005
\mathchardef\Sigma="0006
\mathchardef\Upsilon="0007
\mathchardef\Phi="0008
\mathchardef\Psi="0009
\mathchardef\Omega="000A
\mathchardef\varGamma="0100
\mathchardef\varDelta="0101
\mathchardef\varTheta="0102
\mathchardef\varLambda="0103
\mathchardef\varXi="0104
\mathchardef\varPi="0105
\mathchardef\varSigma="0106
\mathchardef\varUpsilon="0107
\mathchardef\varPhi="0108
\mathchardef\varPsi="0109
\mathchardef\varOmega="010A
%%
%%

%%\catcode`\@=\active
\catcode`\@=12  %% defining '@' as a letter
%%%%%%%%%%%%%%%%%%%%%%%%%%%%%%%%%%%%%%%%%%%%%%%%%%%%%%%%%%%%%%%%%%%%%
\font\tenbf=cmbx10

\font\tenit=cmti10
\font\ninebf=cmbx9
\font\ninerm=cmr9
\font\nineit=cmti9

\font\eightrm=cmr8
\font\eightit=cmti8

\def\section#1{\vskip0pt plus.1\vsize\penalty-250
 \vskip0pt plus-.1\vsize\bigskip\vskip\parskip
 \leftline{\tenbf#1}\nobreak\vglue 5pt}

\def\subsection#1{\medbreak\noindent{\it\ignorespaces
   #1\unskip.\enspace}\ignorespaces}

\outer\def\demo#1{\par\ifdim\lastskip<\smallskipamount\removelastskip
    \smallskip\fi\noindent{\smc\ignorespaces#1\unskip:\enspace}\rm
      \ignorespaces}
\outer\def\enddemo{\par\smallskip}
\def\qed{\hbox{${\vcenter{\vbox{
    \hrule height 0.4pt\hbox{\vrule width 0.4pt height 6pt
    \kern5pt\vrule width 0.4pt}\hrule height 0.4pt}}}$}}

\outer\def\proclaim#1{\medbreak\noindent\smc\ignorespaces
    #1\unskip.\enspace\sl\ignorespaces}
\outer\def\endproclaim{\par\ifdim\lastskip<\medskipamount\removelastskip
  \penalty 55 \fi\medskip\rm}
\outer\def\rmproclaim#1{\medbreak\noindent\smc\ignorespaces
    #1\unskip.\enspace\ignorespaces\rm}
\outer\def\endrmproclaim{\par\ifdim\lastskip<\medskipamount\removelastskip
  \penalty 55 \fi\medskip}

%%% for typesetting the references
\def\references{\vglue12pt plus.1\vsize\penalty-250
 \vskip0pt plus-.1\vsize\bigskip\vskip\parskip
 \message{References}
 \leftline{\tenbf References}\nobreak\vglue 5pt
 \baselineskip=11pt
 \ninerm
 \baselineskip=11pt
 \frenchspacing
}
\def\no#1#2\par{\item{#1.}#2\par}
\def\jr#1{{\nineit#1}}
\def\book#1{{\nineit#1}}
\def\vl#1{{\ninebf#1}}
%%%
%%% Examples:
%%% \no 1 %for a reference number
%%% W. H. Press, B. P. Flannery, S. A. Teukolsky, and W. T. Vetterling,
%%% \book %for a book
%%% {Numerical Recipes:
%%% The Art of Scientific Computing},
%%% (Cambridge Univ. Press, 1986).
%%%
%%% \no 2
%%% L. Trafton,
%%% \jr{J. Comp. Phys.}
%%% \vl %for a volume number
%%% 8
%%% (1971) 64.
%%%
%%% \no 3
%%% R. A. Finkel, in
%%% \book{The Characteristics of Parallel
%%% Algorithms},
%%% eds. L. H. Jamieson, D. B. Gannon, and R. J. Douglass,
%%% (MIT Press, 1987), pp. 21--63.

%%% \, is thin space
%%% \> is medium space
%%% \; is thick space
%%% \! is negative thin space

\def\=def{\; \mathop{=}_{\text{\rm def}} \;}
\def\del{\partial}
\def\res{\, \mathop{\text{\rm res}} \,}

\def\bR{{\bold R}}

\def\L{{\cal L}}
\def\M{{\cal M}}
\def\B{{\cal B}}

\def\F{{\cal F}}
\def\K{{\cal K}}
\def\P{{\cal P}}
\def\Q{{\cal Q}}

\def\Lhat{\hat{\L}}
\def\Mhat{\hat{\M}}

\def\gl{\text{\rm gl}}

%%%%% heading %%%%%%%%%%%%%%%%%%%%%%%%%%%%%%%%%%%%%%%%%%%%%%%%%%%%%%%%
\TagsOnRight
\hsize=5.0truein
\vsize=7.8truein
\parindent=15pt
\nopagenumbers
\baselineskip=10pt

\overfullrule=0pt

\pageno=1

%%% headlines on the first page
\line{\eightrm
Proceedings of the RIMS Research Project 91 ``Infinite Analysis" \hfil
RIMS-814}
\vglue 5pc

\baselineskip=13pt

%%% headline
\headline{\ifnum\pageno=1\hfil\else%
{\ifodd\pageno\rightheadline \else \leftheadline\fi}\fi}

\def\rightheadline{\hfil\eightit
SDiff(2) KP hierarchy
\quad\eightrm\folio}

\def\leftheadline{\eightrm\folio\quad\eightit
Kanehisa Takasaki and Takashi Takebe
\hfil}

\voffset=2\baselineskip

%%% title, author, affiliation %%%%%%%%%%%%%%%%%%%%%%%%%

\centerline{\tenbf SDIFF(2) KP HIERARCHY}
\vglue 24pt

\centerline{\eightrm
KANEHISA TAKASAKI$^1$ and TAKASHI TAKEBE$^2$ }
\baselineskip=12pt

\centerline{\eightit
$^1$ Institute of Mathematics, Yoshida College, Kyoto University}
\baselineskip=10pt

\centerline{\eightit
Yoshida-Nihonmatsu-cho, Sakyo-ku, Kyoto 606, Japan}
\baselineskip=12pt

\centerline{\eightit
$^2$ Department of Mathematics, Faculty of Science, University of Tokyo}
\baselineskip=10pt

\centerline{\eightit
Hongo, Bukyo-ku, Tokyo 606, Japan}
\vglue 20pt
\centerline{Revised version, January 1993}
\vglue 16pt
\vglue 20pt

\centerline{\eightrm ABSTRACT}
{\rightskip=1.5pc
\leftskip=1.5pc
\eightrm\parindent=1pc
An analogue of the KP hierarchy, the SDiff(2) KP hierarchy,
related to the group of area-preserving diffeomorphisms on a
cylinder is proposed.  An improved Lax formalism of the KP
hierarchy is shown to give a prototype of this new hierarchy.
Two important potentials, $S$ and $\tau$, are introduced.
The latter is a counterpart of the tau function of the ordinary
KP hierarchy.  A Riemann-Hilbert problem relative to the group
of area-diffeomorphisms gives a twistor theoretical description
(nonlinear graviton construction) of general solutions.
A special family of solutions related to topological minimal
models are identified in the framework of the Riemann-Hilbert
problem. Further, infinitesimal symmetries of the hierarchy
are constructed. At the level of the tau function, these
symmetries obey anomalous commutation relations, hence
leads to a central extension of the algebra of infinitesimal
area-preserving diffeomorphisms (or of the associated Poisson
algebra).
\vglue12pt}
\baselineskip=13pt
%

%%%%%%%%%%%%%%%%%%%%%%%%%%%%%%%%%%%%%%%%%%%%%%%%%%%%%%%%%%%%%%%%%

\section{1. Introduction}

\noindent
This article presents, with technical details, our recent attempt
to construct an analogue of the KP hierarchy with a different Lie
algebraic structure.  The ordinary KP hierarchy is known to be
characterized by the algebra $\gl(\infty)$ of infinite matrices
[1][2]. Our idea is to replace this algebra by the Lie algebra
of Hamiltonian vector fields (or the Poisson algebra) associated
with area-preserving diffeomorphisms on a surface (in this case, a
cylinder $S^1 \times \bR^1$).  We refer to such groups, in general,
of area-preserving diffeomorphisms as ``SDiff(2)" somewhat
symbolically; ``2" means that we deal with a two-dimensional
manifold on which to consider diffeomorphisms.  Our goal is to
verify that a number of remarkable properties of the ordinary
KP hierarchy persist in this SDiff(2) version.

Our attempt is primarily motivated by recent work of Krichever
[3] on the notion of ``dispersionless Lax equations" and
its application to ``topological minimal models" [4][5].
Krichever's dispersionless Lax equations are a kind of
``quasi-classical" version of ordinary Lax equations for
the KP and generalized KdV equations.  A main characteristic is
that the commutator on the right hand side of Lax equations in the
ordinary sense is now replaced by a Poisson bracket. Despite of
this difference, Krichever pointed out that the notion of the tau
function can be extended to these equations (at least for a class
of special solutions).  This strongly suggests that dispersionless
Lax equations can be treated more systematically along the line of
the Kyoto group [1][2].

Actually, we were first led to this issue from the study of a
different nonlinear equation.
This equation (an SDiff(2) version of Toda field theory) was
first discovered as a 3-d reduction of the 4-d self-dual vacuum
Einstein equation by Boyer and Finley [6] and studied
in more detail by Gegenberg and Das [7].  Since this
is a continuous limit of the ordinary Toda chain, methods of nonlinear
integrable systems have been applied by Golenisheva-Kutuzova and Reiman
[8] and by Saveliev and his coworkers
[9][10]. Bakas [11] and Park [12]
studied the same equation in the context of extended conformal
symmetries ($w_\infty$ algebras).  Park presented a remarkable
observation on hidden symmetries of this equation. Meanwhile,
twistor people [13][14][15][16] arrived at the same equation
from a different direction (``minitwistor theory").
Inspired by these observations, we introduced the notion of the
``SDiff(2) Toda hierarchy" and attempted to apply the approach
of the Kyoto group to this new nonlinear system [17].
In the course of that study, we encountered the work of Krichever
and noticed that his dispersionless Lax equations are very similar
to our SDiff(2) Toda hierarchy.  Therefore we decided to deal with
these two cases in a parallel way, incorporating in particular the
twistor theoretical point of view into Krichever's dispersionless
Lax equations.  This program indeed has turned out to be very
useful.

We should further mention that a prototype of such a twistor theory
of the SDiff(2) KP hierarchy can also be found in Orlov's work [18]
on the KP hierarchy. Actually, we learned Orlov's work through a paper
by Awada and Sin [19], who applied it to $d=1$ string theory.
An essence of Orlov's idea is to enlarge the usual Lax formalism
of the KP hierarchy with a single Lax operator $L$ by adding another
Lax operator $M$. $L$ and $M$ give a ``canonical conjugate
pair" as $[L,M]=1$.  Our twistor theoretical approach to Krichever's
dispersionless Lax equations is based upon a similar pair $\L$ and
$\M$ that are now functions rather than operators, and give a
``classical" canonical conjugate pair for a Poisson bracket as
$\{\L,\M\}=1$.  This provides us with a nice dictionary between
the ordinary and SDiff(2) KP hierarchies.

We first review Orlov's idea in Section 2. The SDiff(2) KP hierarchy
is introduced in Section 3 along with one of our main technical device,
a K\"ahler-like 2-form.  Section 4 is devoted to the notion of the
$S$ function.  This is a key object in Krichever's approach. We do not
actually need the $S$ function because the $(\L,\M)$-pair plays
essentially the same role. It turns out that the $S$ function
is a counterpart of the logarithm of a wave function (a solution of
the linear system) for the KP hierarchy. The notion of tau function is
introduced in Section 5. A twistor theoretical description (which
is a kind of Riemann-Hilbert problem in the SDiff(2) group) of general
solutions is treated in Section 6. This idea is applied to special
solutions related to topological minimal models (Section 7) and to
the construction of infinitesimal symmetries on the space of solutions
(Section 8). In the final stage, infinitesimal symmetries are extended
to the tau function and exhibit, at that level, anomalous commutation
relations.  A central extension of the SDiff(2) algebra thus naturally
emerges. In Section 9, we give a few concluding remarks.

\section{2. $(L,M)$-pair in KP hierarchy}

\noindent
The KP hierarchy describes a set of isospectral deformations of
a first order pseudo differential operator
$$
    L = \del + \sum_{i=1}^\infty u_{i+1}(t)\del^{-i},  \tag 2.1
$$
where $t=(t_1,t_2,\ldots)$ are deformation parameters and
$$
    \del = \del/\del x, \quad x =t_1.                  \tag 2.2
$$
The deformation equations can be written in the so called Lax form as
$$
    \frac{\del L}{\del t_n} = [ B_n, L], \quad
    B_n \=def ( L^n )_{\ge 0},                         \tag 2.3
$$
where $(\quad)_{\ge 0}$ stands for the differential operator part
(i.e., nonnegative powers of $\del$). They give, formally, the
Frobenius integrability conditions of the linear system
$$
    \frac{\del \psi(t,\lambda)}{\del t_n} = B_n \psi(t,\lambda), \quad
    \lambda \psi(t,\lambda) = L \psi(t,\lambda).       \tag 2.4
$$
The system of Lax equations has a zero-curvature representation of
the Zakharov-Shabat form,
$$
    \frac{\del B_m}{\del t_n} - \frac{\del B_n}{\del t_m}
    + [B_m, B_n] = 0.                                  \tag 2.5
$$
It is also known that a zero-th order pseudo-differential operator
$$
    W = 1 + w_1 \del^{-1} + w_2 \del^{-2} + \cdots     \tag 2.6
$$
exists and converts the above equations into the Sato form
$$
    \frac{\del W}{\del t_n} = B_n W - W\del^n,  \quad
    L = W \del W^{-1}.                                 \tag 2.7
$$
Accordingly, the linear system has a solution of the form
$$
    \psi(t,\lambda) = W \exp(\sum_{n=1}^\infty t_n \lambda^n). \tag 2.8
$$
It is this pseudo-differential operator $W$ that encodes the KP
hierarchy into a dynamical system on an infinite dimensional Grassmannian
manifold [1][2][20]. In that respect, one should consider the Lax
operator rather a secondary object.  Actually, the Lax operator $L$
cannot reproduce $W$ uniquely.

Orlov's idea [18] is to introduce the pseudo-differential operator
$$
    M \=def W (\sum_{n=1}^\infty nt_n \del^{n-1}) W^{-1}
      = WxW^{-1} + \sum_{n=2}^\infty nt_n L^{n-1}            \tag 2.9
$$
as a second Lax operator. In fact, $M$ satisfies Lax equations
$$
    \frac{\del M}{\del t_n} = [B_n, M]                       \tag 2.10
$$
and the linear equation
$$
    \frac{\del \psi(t,\lambda)}{\del \lambda} = M \psi(t,\lambda).
                                                             \tag 2.11
$$
Further, $L$ and $M$ obey the canonical commutation relation
$$
    [L, M] = 1.                                              \tag 2.12
$$
With this extended Lax formalism, Orlov pointed out the existence of an
infinite number of infinitesimal symmetries of the KP hierarchy.
These symmetries can be identified with those originating in the
geometry of the infinite dimensional Grassmannian manifold [1][2].
One may thus reorganize the KP hierarchy in terms of the $(L,M)$-pair.

The linear equations for $\psi(t,\lambda)$ yield several interesting
relations.  Note first that Eq. (2.1) has always an inversion formula:
$$
    \del = L + \sum_{i=1}^\infty q_{i+1} L^{-i}.             \tag 2.13
$$
This is a general fact independent of the KP hierarchy. Meanwhile,
because of the linear equation for $L$, one has
$$
    \del \psi(t,\lambda)
    = (\lambda +\sum_{i=1}^\infty q_{i+1}\lambda^{-i})\psi(t,\lambda),
                                                            \tag 2.14
$$
hence
$$
    \frac{\del\log\psi(t,\lambda)}{\del x}
    = \lambda + \sum_{i=1}^\infty q_{i+1}\lambda^{-i}.      \tag 2.15
$$
Similarly, one can expand $M$ and $B_n$ in powers of $L$ as
$$
\align
    M &= \sum_{n=1}^\infty nt_n L^{n-1}
                    +\sum_{i=1}^\infty v_{i+1}L^{-i-1},              \\
    B_n & = L^n + \sum_{i=1}^\infty q_{n,i+1}L^{-i}.       \tag 2.16 \\
\endalign
$$
Then from the linear equations,
$$
\align
    \frac{\del\log\psi(t,\lambda)}{\del\lambda}
       &= \sum_{n=1}^\infty nt_n \lambda^{n-1}
         +\sum_{i=1}^\infty v_{i+1}\lambda^{-i-1},                   \\
    \frac{\del\log\psi(t,\lambda)}{\del t_n}
       &= \lambda^n + \sum_{i=1}^\infty q_{n,i+1}\lambda^{-i}.
                                                           \tag 2.17 \\
\endalign
$$
If one expands $\log\psi(t,\lambda)$ as
$$
    \log\psi(t,\lambda) = \sum_{n=1}^\infty t_n\lambda^n
       +\sum_{i=1}^\infty S_{i+1}\lambda^{-i},             \tag 2.18
$$
the previous Laurent coefficients can be written
$$
    v_{i+1} = -iS_{i+1}, \quad
    q_{i+1} = \frac{\del S_{i+1}}{\del x}, \quad
    q_{n,i+1} = \frac{\del S_{i+1}}{\del t_n}.             \tag 2.19
$$
The coefficients $q_{i+1}$ are related to conservation laws of the
KP hierarchy [21] [22] [23] . %%% cf Takebe, CMP
The other coefficients, too, will have a similar interpretation.
We shall find that all the above relations have some counterpart
in the SDiff(2) KP hierarchy.

\section{3. SDiff(2) KP hierarchy and K\"ahler-like 2-form}

\noindent
Krichever's proposal is to consider a ``quasi-classical" version of
the KP hierarchy replacing $\del$ by $\lambda$ and commutators by
Poisson brackets:
$$
\align
    [\quad,\quad] &\longrightarrow \{\quad,\quad\},              \\
    [\del,x] =1   &\longrightarrow \{\lambda,x\} = 1.    \tag 3.1\\
\endalign
$$
Pseudo-differential operators will then be replaced by Laurent series
of $\lambda$.  According to this prescription, it would be natural
to consider the following system of equations as an analogue of Orlov's
improved Lax formalism of the KP hierarchy.
$$
\align
    & \frac{\del \L}{\del t_n} = \{ \B_n, \L \}, \quad
      \frac{\del \M}{\del t_n} = \{ \B_n, \M \},               \\
    & \{ \L, \M \} = 1,                               \tag 3.2 \\
\endalign
$$
where $\L$ and $\M$ are Laurent series in $\lambda$,
$$
\align
    & \L \=def \lambda + \sum_{i=1}^\infty u_{i+1}\lambda^{-i}, \\
    & \M \=def \sum_{n=1}^\infty nt_n \L^{n-1}
         + \sum_{i=1}^\infty v_{i+1}\L^{-i-1},        \tag 3.3  \\
\endalign
$$
and the $\B$'s are given by the polynomial part (i.e., nonnegative powers
of $\lambda$) of powers of $\L$,
$$
    \B_n \=def ( \L^n )_{\ge 0}.                      \tag 3.4
$$
(The projection $(\quad)_{\ge 0}$ now means the polynomial
part of Laurent series of $\lambda$.)
Finally, $\{\quad,\quad\}$ stands for the Poisson bracket
$$
    \{ F, G \} \=def \frac{\del F}{\del\lambda}\frac{\del G}{\del x}
                    -\frac{\del G}{\del\lambda}\frac{\del F}{\del x}.
                                                     \tag 3.5
$$
We call this system the SDiff(2) KP hierarchy. ``SDiff(2)" now
refers to the structure of a Poisson algebra given by the above
Poisson bracket, which corresponds to the group of area-preserving
diffeomorphisms on a cylinder $S^1 \times \bR^1$.

A number of characteristics of the KP hierarchy, indeed, persist
in this hierarchy. For example, one can prove the following
fact with purely algebraic manipulation (as proven for the
ordinary KP hierarchy [2]).

\proclaim{Proposition 1} The Lax equations for $\L$ are equivalent to the
``zero-curvature equations"
$$
    \frac{\del \B_m}{\del t_n} - \frac{\del \B_n}{\del t_m}
    +\{ \B_m, \B_n \} = 0                           \tag 3.6
$$
and to its ``dual" form
$$
    \frac{\del \B^{-}_m}{\del t_n} - \frac{\del \B^{-}_n}{\del t_m}
    -\{ \B^{-}_m, \B^{-}_n \} = 0,                  \tag 3.7
$$
where
$$
    \B^{-}_n \=def \L^n - \B_n = (\L^n)_{\le -1}    \tag 3.8
$$
and $(\quad)_{\le -1}$ stands for the negative power part of Laurent
series of $\lambda$.
\endproclaim

\noindent
A crucial difference, however, is that there is no direct counterpart
of the $W$ operator. The successful understandings of the KP hierarchy
[1][2][20] are all based upon the use of $W$.
We need something new in place of $W$ for the study of the SDiff(2)
version.

We already know equations of the above type in the analysis of the
self-dual vacuum Einstein equation and hyper-K\"ahler geometry [24].
Actually, they are slightly different in the sense that the spectral
variable $\lambda$ therein is merely a parameter. In the above
setting, one has to treat $\lambda$ as a true variable that enters
into the definition of the Poisson bracket along with $x$.
Apart from this difference, both situations are almost the same.
In the study of the vacuum Einstein equation (as well as its
hyper-K\"ahler version), a K\"ahler-like 2-form and associated
``Darboux coordinates" play a central role.  We now show that
the SDiff(2) KP hierarchy has a similar structure.  Let $\omega$
be a 2-form given by
$$
    \omega \=def \sum_{n=1}^\infty d\B_n \wedge dt_n
           = d\lambda \wedge dx + \sum_{n=2}^\infty d\B_n \wedge dt_n,
                                                        \tag 3.9
$$
where ``$d$" now stands for total differentiation in both $t$ and
$\lambda$. From the definition, obviously, $\omega$ is a closed form,
$$
    d\omega = 0.                                        \tag 3.10
$$
The zero-curvature equations for $\B_n$ can be cast into a compact form
as
$$
    \omega \wedge \omega = 0.                           \tag 3.11
$$
These two relations ensure the existence of ``Darboux coordinates"
$\P$ and $\Q$ (functions of $t$ and $\lambda$) such that
$$
    \omega = d\P \wedge d\Q.                            \tag 3.12
$$
(In the case of the self-dual vacuum Einstein equation, $\lambda$ is
considered a constant under the total differentiation.)  Actually,
$\L$ and $\M$ give (and are characterized as) such a pair of
functions:

\proclaim{Proposition 2} The SDiff(2) KP hierarchy is equivalent to
the exterior differential equation
$$
    \omega = d\L \wedge d\M.                            \tag 3.13
$$
\endproclaim

\demo{Proof} We only show the derivation of the Lax system from the exterior
differential equation; the converse can be checked by simply tracing back
the following reasoning.  Expanding both sides of the exterior differential
equation as linear combination of $d\lambda \wedge dt_n$ and
$dt_m \wedge dt_n$ give rise to an infinite set of partial differential
equations.  From coefficients of $d\lambda \wedge dx$ ($x=t_1$), one has
$$
    1 = \frac{\del \L}{\del \lambda}\frac{\del \M}{\del x}
       -\frac{\del \L}{\del x}\frac{\del \M}{\del \lambda}
      = \{ \L, \M \}.                                   \tag 3.14
$$
Similarly, from coefficients of $d\lambda \wedge dt_n$ and
$dx \wedge dt_n$, respectively,
$$
\align
    & \frac{\del \B_n}{\del \lambda}
      = \frac{\del \L}{\del \lambda}\frac{\del \M}{\del t_n}
       -\frac{\del \M}{\del \lambda}\frac{\del \L}{\del t_n},
                                                                   \\
    & \frac{\del \B_n}{\del x}
      = \frac{\del \L}{\del x}\frac{\del \M}{\del t_n}
       -\frac{\del \M}{\del x}\frac{\del \L}{\del t_n}.  \tag 3.15
                                                                   \\
\endalign
$$
One can easily solve these relations for $\del \L/\del t_n$
and $\del \M / \del t_n$ because the coefficient matrix is
unimodular, $\{\L,\M\}=1$, as we have just deduced above.
The result is:
$$
\align
    & \frac{\del \L}{\del t_n}
      = \frac{\del \B_n}{\del \lambda}\frac{\del \L}{\del x}
       -\frac{\del \B_n}{\del x}\frac{\del \L}{\del \lambda}
      = \{ \B_n, \L \},
                                                                   \\
    & \frac{\del \M}{\del t_n}
      = \frac{\del \B_n}{\del \lambda}\frac{\del \M}{\del x}
       -\frac{\del \B_n}{\del x}\frac{\del \M}{\del \lambda}
      = \{ \B_n, \M \},                                  \tag 3.16
                                                                   \\
\endalign
$$
This completes the proof.
\qed
\enddemo

\section{4. $S$ function}

\noindent
Note that the fundamental relation
$$
    \omega = d\L \wedge d\M                                   \tag 4.1
$$
can be rewritten
$$
    d\left( \M d\L + \sum_{n=1}^\infty \B_n dt_n \right) = 0. \tag 4.2
$$
This implies the existence of a function $S$ such that
$$
    dS = \M d\L + \sum_{n=1}^\infty \B_n dt_n,                \tag 4.3
$$
or, equivalently,
$$
\align
    & \M = \left(\frac{\del S}{\del \L}\right)
                       _{t\ \text{\rm fixed}},                \tag 4.4 \\
    & \B_n = \left(\frac{\del S}{\del t_n}\right)
                       _{\L,t_m(m\not=n)\ \text{\rm fixed}}.  \tag 4.5 \\
\endalign
$$
This potential $S$ is introduced by Krichever [3].
In his formulation, $\M$ is implicit in various calculations and
the $S$ function plays rather a central role.

Actually, we have the following explicit realization of the $S$
function in terms of the Laurent coefficients $v_{i+1}$ of $\M$.

\proclaim{Proposition 3} $S$ is given by
$$
    S = \sum_{i=1}^\infty t_i \L^i + \sum_{i=1}^\infty S_{i+1}\L^{-i},
    \quad S_{i+1} = - \frac{1}{i}v_{i+1}.                    \tag 4.6
$$
In particular, $\B_n$ can be written explicitly in terms of the Laurent
coefficients $S_{i+1}$ as
$$
    \B_n = \L^n + \sum_{i=1}^\infty \frac{\del S_{i+1}}{\del t_n}\L^{-i}.
                                                             \tag 4.7
$$
\endproclaim

It is amusing to compare this result with the KP hierarchy (Section 2).
One will immediately find that the $S$ function plays the same role as
the logarithm of $\psi$. Note, in particular, that Eq. (4.7) for
$n=1$ becomes
$$
    \lambda = \L + \sum_{i=1}^\infty \frac{\del S_{i+1}}{\del x}\L^{-i},
                                                             \tag 4.8
$$
hence gives an explicit inversion formula
$$
    \lambda = \L + \sum_{i=1}^\infty q_{i+1}\L^{-i}          \tag 4.9
$$
of the Laurent expansion of $\L$ in $\lambda$. These coefficients
$q_{i+1}$ are known in the theory of singularities as
``flat coordinates" [25]. It is thus natural that the notion of
flat coordinates plays an important role in topological minimal
models [4][5] (see Section 7).

Let us give a proof of the above result.  To this end, we use the notion
of formal residue of 1-forms:
$$
    \res \sum a_n \lambda^n d\lambda \=def a_{-1}.           \tag 4.10
$$
and the following basic properties without proofs.

\proclaim{Lemma A} For any (formal) Laurent series $F$ and $G$ of $\lambda$,
$$
\align
    & \res d_\lambda F = 0,                                  \tag 4.11 \\
    & \res F d_\lambda G = -\res G d_\lambda F,              \tag 4.12 \\
    & \res F d_\lambda G = \res (F_{\ge 0}) d_\lambda (G_{\le -1})
                         + \res (F_{\le -1}) d_\lambda (G_{\ge 0}),
                                                             \tag 4.13 \\
\endalign
$$
where ``$d_\lambda$" stands for total differentiation with respect to
$\lambda$.
\endproclaim

\proclaim{Lemma B} For any integer $n$,
$$
    \res \L^n d_\lambda \L = \delta_{n,-1}.                  \tag 4.14
$$
\endproclaim

\noindent
Bearing these observations in mind, we first prove:

\proclaim{Proposition 4} The $t$-derivatives of $v_{i+1}$ are given by
$$
    \frac{\del v_{i+1}}{\del t_n} = \res \L^i d_\lambda\B_n.  \tag 4.15
$$
\endproclaim

\demo{Proof} By the chain rule of differentiation,
$$
    \frac{\del\M}{\del t_n}
    = \left( \frac{\del\M}{\del\L} \right)_{t,v\;\text{\rm fixed}}
        \frac{\del\L}{\del t_n}
     +n\L^{n-1}
     +\sum_{i=1}^\infty \frac{\del v_{i+1}}{\del t_n}\L^{-i-1},
                                                              \tag 4.16
$$
where ``$t,v\;\text{\rm fixed}$" means that $\M$ is differentiated
with respect to $\L$ while the Laurent coefficients $t_n$ and $v_i$
being fixed. In other words,
$$
    \left( \frac{\del\M}{\del\L} \right)_{t,v\;\text{\rm fixed}}
    = \sum_{n=1}^\infty n(n-1)t_n\L^{n-2}
     -\sum_{i=1}^\infty (i+1)v_{i+1}\L^{-i-2},
$$
though one does not need this explicit expansion.  Therefore, using
Lemma B as well, one has
$$
    \frac{\del v_{i+1}}{\del t_n}
    = \res
        \L^i \left[
          \frac{\del\M}{\del t_n}
         -\left( \frac{\del\M}{\del\L} \right)_{t,v\;\text{\rm fixed}}
            \frac{\del\L}{\del t_n}
        \right]d_\lambda\L.                                  \tag 4.17
$$
The ``$[\ldots]$" part times $d_\lambda\L$ in the last formula can be
calculated as:
$$
\align
    [\cdots]d_\lambda\L
    &=  \frac{\del\M}{\del t_n}d_\lambda\L
       -\frac{\del\L}{\del t_n}d_\lambda\M
     =  \{ \B_n, \M \}d_\lambda\L -\{ \B_n, \L \}d_\lambda\M
                                                               \\
    &= \left(
          \frac{\del\B_n}{\del\lambda}\frac{\del\M}{\del x}
         -\frac{\del\B_n}{\del x}\frac{\del\M}{\del\lambda}
       \right)
         \frac{\del\L}{\del\lambda} d\lambda
      -\left(
          \frac{\del\B_n}{\del\lambda}\frac{\del\L}{\del x}
         -\frac{\del\B_n}{\del x}\frac{\del\L}{\del\lambda}
       \right)
         \frac{\del\M}{\del\lambda} d\lambda
                                                               \\
    &= \left(
          \frac{\del\L}{\del\lambda}\frac{\del\M}{\del x}
         -\frac{\del\L}{\del x}\frac{\del\M}{\del\lambda}
       \right)
       \frac{\del\B_n}{\del\lambda} d\lambda
     = d_\lambda \B_n.                               \tag 4.18 \\
\endalign
$$
(We have also used the canonical Poisson relation of $\L$ and
$\M$.)
\qed
\enddemo

\demo{Proof of Proposition 3} Eq. (4.4) is obviously satisfied
because of the construction. One has to prove Eq. (4.5) or,
equivalently, Eq. (4.7). Use the previous lemmas to continue
the right hand side of Eq. (4.15) as
$$
    \frac{\del v_{i+1}}{\del t_n}
    = -\res \B_n d_\lambda(\L^i)
    = -i \res \B_n \L^{i-1} d_\lambda\L.               \tag 4.19
$$
{}From these relations for $i \ge 1$ (and Lemma B), $\B_n$ turns out to
have a Laurent expansion as
$$
    \B_n = \sum_{m\ge 0} b_{nm}\L^m
          -\sum_{i=1}^\infty \frac{1}{i}
                             \frac{\del v_{i+1}}{\del t_n}\L^{-i}
                                                        \tag 4.20
$$
with yet undetermined coefficients $b_{nm}$. The $(\quad)_{\ge 0}$
part of both hand sides then gives rise to such a relation as
$$
    \B_n = \sum_{m \ge 0} b_{nm} \B_m,                  \tag 4.21
$$
however $B_m$'s should be linearly independent polynomials of
$\lambda$, hence
$$
    b_{nm} = \delta_{nm}.                               \tag 4.22
$$
Since $S_{i+1}=-v_{i+1}/i$, the last expression of $\B_n$ gives Eq. (4.7).
\qed\enddemo

We conclude this section with a comment on the missing coefficients
$u_1$ and $v_1$ in the definition of the hierarchy.  We have excluded
these terms because they are absent in the ordinary KP hierarchy.
Actually, we could have included them in $\L$ and $\M$, but it turns
out that they can be absorbed into redefinition of $\lambda$ and $\M$.
This is due to the following fact.

\proclaim{Proposition 5} Suppose that $u_1$ and $v_1$ are inserted
as
$$
    \L = \lambda + \sum_{i=0}^\infty u_{i+1}\lambda^{-i}, \quad
    \M = \sum_{n=1}^\infty nt_n \lambda^{n-1}
        +\sum_{i=0}^\infty v_{i+1} \lambda^{-i-1}             \tag 4.23
$$
and the same Lax equations (including the $n=1$ case) and the canonical
Poisson relation are satisfied.  Then $u_1$ and $v_1$ become constants.
\endproclaim

\demo{Proof}
Consider first the Lax equation for $\L$. The right hand side can be
evaluated as
$$
\align
    & \{ \B_n, \L \}
      = -\{ (\L^n)_{\le -1}, \L \}                             \\
    & = \{ O(\lambda^{-1}), \lambda + u_1 + O(\lambda^{-1}) \}
      = O(\lambda^{-1}).                           \tag 4.24   \\
\endalign
$$
hence from the $\lambda^0$-part,
$$
    \frac{\del u_1}{\del t_n} = 0.                      \tag 4.25
$$
To prove that $v_1$ is a constant, we note that Eq. (4.15) is also
valid for $i=0$. Accordingly,
$$
    \frac{\del v_1}{\del t_n} = \res d_\lambda \B_n =0. \tag 4.26
$$
This completes the proof.
\qed\enddemo

\noindent
Consequently, $u_1$ and $v_1$ can be absorbed by redefinition of
$\lambda$ and $\M$ as
$$
\align
    \lambda + u_1   & \longrightarrow \lambda,                \\
    \M - v_1\L^{-1} & \longrightarrow \M.           \tag 4.27 \\
\endalign
$$
With this prescription, one may always assume
$$
    u_1=0, \quad v_1=0                            \tag 4.28
$$
without loosing generality.  Later, however, we shall have to
relax the second condition because the nonlinear graviton
construction (see Section 6) can, in general, generate a nonzero
$v_1$ term.

\section{5. Tau function}

\noindent
We define the tau function $\tau$ of the SDiff(2) KP hierarchy
by the equations
$$
    \frac{\del \log\tau}{\del t_n} = v_{n+1}, \quad n=1,2,\ldots.
                                                          \tag 5.1
$$
This is due to the following basic fact.

\proclaim{Proposition 6} The functions $v_{n+1}$ on the right hand side
satisfy the integrability condition
$$
    \frac{\del v_{m+1}}{\del t_n} = \frac{\del v_{n+1}}{\del t_m},
    \quad m,n=1,2,\ldots.                                 \tag 5.2
$$
\endproclaim

\demo{Proof}
{}From Eq. (4.15) and Lemma A,
$$
    \frac{\del v_{m+1}}{\del t_n}
    = \res (\L^m)_{\le -1} d_\lambda (\L^n)_{\ge 0}.      \tag 5.3
$$
Taking difference after interchanging $m \leftrightarrow n$ and
using Lemmas A and B, one has
$$
\align
    \frac{\del v_{m+1}}{\del t_n} - \frac{\del v_{n+1}}{\del t_m}
    &= \res[(\L^m)_{\le -1} d_\lambda(\L^n)_{\ge 0}]
      -\res[(\L^n)_{\le -1} d_\lambda(\L^m)_{\ge 0}]             \\
    &= \res[(\L^m)_{\le -1} d_\lambda(\L^n)_{\ge 0}]
      +\res[(\L^m)_{\ge 0}  d_\lambda(\L^n)_{\le -1}]            \\
    &= \res[\L^m d_\lambda \L^n ]                                \\
    &= n\delta_{m+n,0},                                   \tag 5.4
                                                                 \\
\endalign
$$
which vanishes because $m$ and $n$ are now positive integers.
\qed\enddemo

The tau function contains all information of the hierarchy. In fact,
we can reproduce $u_i$ and $v_i$ (hence $\L$ and $\M$) from the
tau function. This can be seen as follows. First, from the construction,
the $v$'s can be written
$$
    v_{n+1} = \frac{\del \log\tau}{\del t_n}.              \tag 5.5
$$
To obtain a similar expression for the $u$'s, recall Eqs. (4.6)-(4.9).
{}From these relations, one can see that $q_{n+1}$ can be written
$$
    q_{n+1} = -\frac{1}{n} \frac{\del v_{n+1}}{\del x}
            = -\frac{1}{n} \frac{\del^2 \log\tau}{\del t_n \del x}.
                                                           \tag 5.6
$$
This means that $u_{n+1}$ is a differential polynomial of $\log\tau$,
because the $u_{n+1}$'s and the $q_{n+1}$'s are connected by an
invertible polynomial relation. Actually, they are linked in a more
explicit form by residue formulas:
$$
\align
    & u_{n+1} = -\frac{1}{n} \res \lambda^n d_\lambda \L,           \\
    & q_{n+1} = -\frac{1}{n} \res \L^n d\lambda.           \tag 5.7 \\
\endalign
$$
Thus the SDiff(2) KP hierarchy can be, in principle, rewritten as a
system of differential equations for the tau function $\tau$.
In the case of the ordinary KP hierarchy [1][2], this leads to
the celebrated Hirota bilinear equations.  No similar expression
has been discovered for the SDiff(2) version.

It is instructive to compare our
definition of the tau function with the case of the ordinary KP
hierarchy. To distinguish between the ordinary KP hierarchy and the
SDiff(2) version, we now put superscript ``$^{KP}$" for the
ordinary KP hierarchy as $\tau^{KP}$, $u^{KP}_n$, $v^{KP}_n$,
etc., whereas $\tau$, $u_n$, $v_n$, etc. stand for their SDiff(2)
counterparts. In the KP hierarchy, $\psi=\psi(t,\lambda)$ is linked
with the tau function as [2]
$$
\align
    \log\psi
    &= \sum_{n=1}^\infty t_n\lambda^n
      +\log\dfrac{ \tau^{KP}(t_1-\frac{1}{\lambda},
                        t_2-\frac{1}{2\lambda^2},\ldots) }
                 { \tau^{KP}(t_1,t_2,\ldots) }
                                                                 \\
    &= \sum_{n=1}^\infty t_n\lambda^n
      +\left[ \exp\left( -\sum_{n=1}^\infty
                   \frac{1}{n\lambda^n} \frac{\del}{\del t_n}
              \right) - 1
       \right]
       \log\tau^{KP}(t)
                                                                 \\
    &= \sum_{n=1}^\infty t_n\lambda^n
       +\sum_{N=1}^\infty
          \frac{1}{N!}
          \left(-\sum_{n=1}^\infty
                   \frac{1}{n\lambda^n} \frac{\del}{\del t_n}
          \right)^N
          \log\tau^{KP}(t).                           \tag 5.8   \\
\endalign
$$
We have seen in the previous section that the $S$ function of the
SDiff(2) KP hierarchy should correspond to $\log\psi$ of the
KP hierarchy. In fact, we now have
$$
\align
    S &= \sum_{n=1}^\infty t_n\L^n
        +\sum_{n=1}^\infty S_{n+1}\L^{-n}
                                                                 \\
      &= \sum_{n=1}^\infty t_n\L^n
        -\sum_{n=1}^\infty
            \frac{1}{n\L^n} \frac{\del}{\del t_n}\log\tau(t),
                                                      \tag 5.9   \\
\endalign
$$
which is very similar to the above formula except that

\item{$\bullet$}
only the $N=1$ term is retained, and
\item{$\bullet$}
$\lambda$ is replaced by $\L$.

\section{6. Riemann-Hilbert problem in SDiff(2) group}

\noindent
The nonlinear graviton construction of Penrose [26]
generates, in principle, all (both local and global) solutions
of the self-dual vacuum Einstein equation and its hyper-K\"ahler
version [27].  This method can be extended to
the SDiff(2) KP hierarchy. A key step is to solve the functional
equation
$$
\align
    & f(\L,\M)_{\le -1} = 0,                   \\
    & g(\L,\M)_{\le -1} = 0,          \tag 6.1 \\
\endalign
$$
where $f=f(\lambda,x)$ and $g=g(\lambda,x)$ are arbitrary holomorphic
functions defined in a neighborhood of $\lambda=\infty$ except at
$\lambda=\infty$ itself, and required to satisfy the canonical Poisson
relation
$$
    \{ f(\lambda,x), g(\lambda,x) \} = 1.  \tag 6.2
$$
Thus the pair $(f,g)$ may be thought of as an area-preserving
diffeomorphism. (For simplicity, we do not specify the domain
where $x$ is supposed to take values. That depends on the situation.)
One may rewrite the functional equations
as
$$
\align
    & f(\L, \M) = \Lhat,                      \\
    & g(\L, \M) = \Mhat,             \tag 6.3 \\
\endalign
$$
where $\Lhat$ and $\Mhat$ are another set of unknown functions and
required to have Laurent expansion in $\lambda$ with only nonnegative
powers,
$$
    \Lhat = \Lhat_{\ge 0}, \quad \Mhat = \Mhat_{\ge 0}. \tag 6.4
$$
In the latter expression, the functional equations look more like
a ``Riemann-Hilbert problem" as we know for the case of the
self-dual vacuum Einstein equation and hyper-K\"ahler geometry
[26][27].

We now assume that the above ``Riemann-Hilbert problem" has a
unique solution with $\L$ and $\M$ in the form given in the
definition of the SDiff(2) KP hierarchy. This is indeed ensured,
as Penrose observed in the case of the self-dual vacuum Einstein
equation, if $(f,g)$ is sufficiently close to the trivial one
$(f,g)=(\lambda,x)$.

\proclaim{Proposition 7} Such a solution $(\L,\M)$ of the functional
equations gives a solution of the SDiff(2) KP hierarchy.
\endproclaim

\demo{Proof} We first derive the canonical Poisson relation.
By differentiating Eqs. (6.3) with respect to $\lambda$ and $x$,
$$
    \pmatrix \dfrac{\del f(\L,\M)}{\del\L} & \dfrac{\del f(\L,\M)}{\del\M} \\
             \dfrac{\del g(\L,\M)}{\del\L} & \dfrac{\del g(\L,\M)}{\del\M} \\
    \endpmatrix
    \pmatrix \dfrac{\del\L}{\del\lambda} & \dfrac{\del\L}{\del x} \\
             \dfrac{\del\M}{\del\lambda} & \dfrac{\del\M}{\del x} \\
    \endpmatrix
  = \pmatrix \dfrac{\del\Lhat}{\del\lambda} & \dfrac{\del\Lhat}{\del x} \\
             \dfrac{\del\Mhat}{\del\lambda} & \dfrac{\del\Mhat}{\del x} \\
    \endpmatrix.                                               \tag 6.5
$$
Since the first matrix on the left hand side is unimodular because of
Eq. (6.2), the determinants of both hand sides give
$$
    \{ \L, \M \} = \{ \Lhat, \Mhat \}.                         \tag 6.6
$$
Following the idea of the proof of Proposition 4, one can calculate
the left hand side as
$$
\align
    &\{ \L, \M \}
        = \frac{\del\L}{\del\lambda}\frac{\del\M}{\del x}
         -\frac{\del\M}{\del\lambda}\frac{\del\L}{\del x}    \\
    &= \frac{\del\L}{\del\lambda}
       \left[
          \left(\frac{\del\M}{\del\L}\right)_{t,v\;\text{\rm fixed}}
           \frac{\del\L}{\del x}
          +1
          +\sum_{i=1}^\infty \frac{\del v_{i+1}}{\del x}\L^{-i}
       \right]
      -\frac{\del\L}{\del x}
        \left(\frac{\del\M}{\del\L}\right)_{t,v\;\text{\rm fixed}}
         \frac{\del\L}{\del\lambda},
\endalign
$$
but terms containing $(\del\M/\del\L)_{t,v\;\text{\rm fixed}}$
in the last line cancel, hence
$$
    = 1 + (\text{\rm negative powers of }\lambda).            \tag 6.7
$$
Meanwhile, Laurent expansion of $\{ \Lhat, \Mhat \}$ contains
only nonnegative powers of $\lambda$.  Therefore strictly negative
(as well as) powers of $\lambda$ in the last line should be absent,
thus
$$
    \{ \L, \M \} = \{ \Lhat, \Mhat \} = 1.                    \tag 6.8
$$
This gives the canonical Poisson commutation that we have sought for.
We now show that the Lax equations for $\L$ and $\M$ are indeed satisfied.
Differentiating Eqs. (6.3) now with respect to $t_n$ gives
$$
  \pmatrix \dfrac{\del f(\L,\M)}{\del\L} & \dfrac{\del f(\L,\M)}{\del\M} \\
           \dfrac{\del g(\L,\M)}{\del\L} & \dfrac{\del g(\L,\M)}{\del\M} \\
  \endpmatrix
  \pmatrix \dfrac{\del\L}{\del t_n} \\
           \dfrac{\del\M}{\del t_n} \\
  \endpmatrix
  = \pmatrix \dfrac{\del\Lhat}{\del t_n} \\
             \dfrac{\del\Mhat}{\del t_n} \\
    \endpmatrix.                                               \tag 6.9
$$
Combining Eqs. (6.5) and (6.9), one can eliminate the derivative matrix
of $(f,g)$ by $(\L,\M)$ and obtain the matrix relation
$$
   \left(\matrix
            \dfrac{\del\L}{\del\lambda} & \dfrac{\del\L}{\del x} \\
            \dfrac{\del\M}{\del\lambda} & \dfrac{\del\M}{\del x} \\
   \endmatrix\right)^{-1}
   \left(\matrix  \dfrac{\del\L}{\del t_n} \\
                  \dfrac{\del\M}{\del t_n} \\
   \endmatrix\right)
 = \left(\matrix
            \dfrac{\del\Lhat}{\del\lambda} & \dfrac{\del\Lhat}{\del x} \\
            \dfrac{\del\Mhat}{\del\lambda} & \dfrac{\del\Mhat}{\del x} \\
   \endmatrix\right)^{-1}
   \left(\matrix \dfrac{\del\Lhat}{\del t_n} \\
                 \dfrac{\del\Mhat}{\del t_n} \\
   \endmatrix\right).                                   \tag 6.10
$$
Since the two $2 \times 2$ matrices on both sides are unimodular
because of Eq. (6.8), the inverse can also be written explicitly.
In components, thus, the above matrix relation gives:
$$
\align
     \frac{\del\M}{\del x}\frac{\del\L}{\del t_n}
    -\frac{\del\L}{\del x}\frac{\del\M}{\del t_n}
  &= \frac{\del\Mhat}{\del x}\frac{\del\Lhat}{\del t_n}
    -\frac{\del\Lhat}{\del x}\frac{\del\Mhat}{\del t_n},
                                                                 \\
     \frac{\del\M}{\del \lambda}\frac{\del\L}{\del t_n}
    -\frac{\del\L}{\del \lambda}\frac{\del\M}{\del t_n}
  &= \frac{\del\Mhat}{\del \lambda}\frac{\del\Lhat}{\del t_n}
    -\frac{\del\Lhat}{\del \lambda}\frac{\del\Mhat}{\del t_n}.   \tag 6.11
                                                                 \\
\endalign
$$
The left hand side of Eqs. (6.11) can be calculated just as we
have done above for derivatives in $(x,\lambda)$.
For the first equation of (6.11),
$$
\align
  & \frac{\del\M}{\del x}\frac{\del\L}{\del t_n}
    -\frac{\del\L}{\del x}\frac{\del\M}{\del t_n}              \\
  &= \left[
        \left(\frac{\del\M}{\del\L}\right)_{t,v\;\text{\rm fixed}}
          \frac{\del\L}{\del x}
       +1
       +\sum_{i=1}^\infty \frac{\del v_{i+1}}{\del x} \L^{-i}
     \right]
     \frac{\del\L}{\del t_n}                                   \\
  &\phantom{=}
    -\frac{\del\L}{\del x}
     \left[
        \left(\frac{\del\M}{\del\L}\right)_{t,v\;\text{\rm fixed}}
          \frac{\del\L}{\del t_n}
       +n\L^{n-1}
       +\sum_{i=1}^\infty \frac{\del v_{i+1}}{\del t_n} \L^{-i}
     \right],                                                  \tag 6.12
                                                               \\
\endalign
$$
and terms containing $(\del\M/\del\L)_{t,v\;\text{\rm fixed}}$
cancel. The rest can be easily evaluated. Thus,
$$
      \frac{\del\M}{\del x}\frac{\del\L}{\del t_n}
     -\frac{\del\L}{\del x}\frac{\del\M}{\del t_n}
   = -\frac{\del(\L^n)_{\ge 0}}{\del x}
     +(\text{\rm negative powers of }\lambda).                 \tag 6.13
$$
Similarly,
$$
      \frac{\del\M}{\del\lambda}\frac{\del\L}{\del t_n}
     -\frac{\del\L}{\del\lambda}\frac{\del\M}{\del t_n}
   = -\frac{\del(\L^n)_{\ge 0}}{\del \lambda}
     +(\text{\rm negative powers of }\lambda).                 \tag 6.14
$$
The right hand side of Eqs. (6.11), meanwhile, have Laurent
expansion with only nonnegative powers of $\lambda$.
Therefore only nonnegative powers of $\lambda$ should
survive, thus
$$
\align
 &  \frac{\del\M}{\del x}\frac{\del\L}{\del t_n}
   -\frac{\del\L}{\del x}\frac{\del\M}{\del t_n}
   = -\frac{\del(\L^n)_{\ge 0}}{\del x}
   = -\frac{\del\B_n}{\del x}
                                                                      \\
 &  \frac{\del\M}{\del\lambda}\frac{\del\L}{\del t_n}
   -\frac{\del\L}{\del\lambda}\frac{\del\M}{\del t_n}
   = -\frac{\del(\L^n)_{\ge 0}}{\del \lambda}
   = -\frac{\del\B_n}{\del \lambda}                         \tag 6.15 \\
\endalign
$$
These equations can be readily solved (again due to the unimodular
property, (6.8), of the coefficient matrix):
$$
\align
    & \frac{\del\L}{\del t_n}
      = -\frac{\del\L}{\del\lambda}\frac{\del\B_n}{\del x}
        +\frac{\del\M}{\del x}     \frac{\del\B_n}{\del\lambda}
      = \{ \B_n, \L \},
                                                                      \\
    & \frac{\del\M}{\del t_n}
      = -\frac{\del\M}{\del\lambda}\frac{\del\B_n}{\del x}
        +\frac{\del\M}{\del x}     \frac{\del\B_n}{\del\lambda}
      = \{ \B_n, \M \}.                                    \tag 6.16  \\
\endalign
$$
One can thus derive the Lax equations as well.
\qed
\enddemo

We thus anyway have a solution scheme for general solutions of the
SDiff(2) KP hierarchy.  Unfortunately, finding an explicit form of the
solution of the functional equations is a very hard problem at this
moment. The case of the self-dual vacuum Einstein equation and its
hyper-K\"ahler version is slightly simpler because $\lambda$ therein
is just a parameter unlike the present setting; nevertheless,
explicitly solvable cases are very limited [27]
[28][29].  Very few is known for the SDiff(2) version.
We now turn to this issue.

\section{7. Reductions and special solutions}

\subsection{7.1. Krichever's dispersionless Lax equations}
In Krichever's original formulation [3],
``dispersionless Lax equations" are Lax equations
$$
    \frac{\del \P}{\del t_n} = \{ \B_n, \P \}               \tag 7.1
$$
of a polynomial
$$
    \P \=def \lambda^N + p_2 \lambda^{N-2} + \cdots + p_N  \tag 7.2
$$
rather than an infinite Laurent series like the $\L$. The $\B_n$'s
are now given by
$$
    \B_n = (\P^{n/N})_{\ge 0}.                             \tag 7.3
$$
Obviously, the $N$-th root
$$
   \L = \P^{1/N}                                           \tag 7.4
$$
of $\L$ satisfies the Lax equations of the SDiff(2) KP hierarchy,
and conversely, the above situation is characterized by the condition
$$
    (\L^N)_{\le -1} = 0.                                 \tag 7.5
$$
In the case of the ordinary KP hierarchy [1][2], this amounts
a reduction of generalized KdV type (the KdV equation for $N=2$,
the Boussinesq for $N=3$, etc.).  After the terminology therein,
one may call this reduction ``$N$-reduction."

Eqs. (7.1) give rise to a system of evolution equations
for $p_2$, $\ldots$, $p_N$,
$$
    \frac{\del p_i}{\del t_n}
    = \sum_{j=2}^N f_{ijn}(p_2,\ldots,p_N)\frac{\del p_j}{\del x},
                                                         \tag 7.6
$$
where $f_{ijn}(p_2,\ldots,p_N)$ are polynomials of $p_2$, $\ldots$, $p_N$.
These equations fall into a family of equations of ``hydrodynamic
type" studied by Krichever, Novikov and Tsarev [30][31][32].
Krichever [3] pointed out a link of the above dispersionless
Lax equations with topological minimal models [4][5]. Our goal
in this section is to give an interpretation of his observation
in our framework.

\subsection{7.2. Hodograph transformation method for $N=2$}
According to Tsarev [31], the traditional method of ``hodograph
transformation" can be extended to these equations. We now
illustrate this method in the $N=2$ case.

It is convenient to write $\P$ as
$$
    \P = \lambda^2 + 2p, \quad p = p_2/2.                \tag 7.7
$$
In this case, only ``odd" flows associated with $(t_3,t_5,\ldots)$
are nontrivial. For ``even" flows, we have $\B_{2n}=\L^{2n}$ for
$n=1,2,\ldots$, therefore
$$
\align
    \frac{\del\L}{\del t_{2n}} &= \{ \L^{2n}, \L \} = 0, \tag 7.8
                                                                  \\
    \frac{\del\M}{\del t_{2n}} &= \{ \L^{2n}, \M \} = 2n\L^{2n-1},
                                                         \tag 7.9 \\
\endalign
$$
which shows that $u_i$ and $v_i$ are independent of $t_{2n}$.
To describe odd flows, we introduce a set of polynomials of $p$:
$$
    r_n(p) \=def \res \P^{n+1/2}d\lambda.               \tag 7.10
$$
They correspond to the Gelfand-Dikii resolvent functionals of the
KdV equation [33]. In the present setting, the fractional powers
of $\P$ can be calculated by means of the binomial expansion
of $\P^{n+1/2} = (\lambda^2 + 2p)^{n+1/2}$ as:
$$
    r_n(p) = \left( {{n+\frac{1}{2}}\atop{n+1}} \right) (2p)^{n+1}
           = \frac{(2n+1)!!}{(n+1)!}p^{n+1}.              \tag 7.11
$$
They obey a set of recursion relations just as the Gelfand-Dikii
resolvent functionals:
$$
    \frac{\del r_n(p)}{\del p} = (2n+1)r_{n-1}(p).        \tag 7.12
$$
Now the right hand side of the Lax equations for odd flows can be
written
$$
    \{ (\P^{n+1/2})_{\ge 0}, \P \}
    = - \{ (\P^{n+1/2})_{\le -1}, \P \}
      = 2\frac{r_n(p)}{\del x} + O(\lambda^{-1}).         \tag 7.13
$$
Actually, the left hand side of Eq. (7.13) should be a polynomial of
$\lambda$, hence the remainder term $O(\lambda^{-1})$ on the right
hand side must be absent. Thus the Lax equations can be reduced to
evolution equations of $p$:
$$
    \frac{\del p}{\del t_{2n+1}}
    = \frac{\del r_n(p)}{\del x}
    = (2n+1)r_{n-1}(p) \frac{\del p}{\del x}.            \tag 7.14
$$
The $t_3$-flow gives the ``dispersionless KdV equation"
$$
    \frac{\del p}{\del t_3}
    = \frac{\del}{\del x}(\frac{3}{2}p^2)
    = 3p\frac{\del p}{\del x}.                           \tag 7.15
$$
The method of ``generalized hodograph transformations" [31]
gives a general solution $p=p(x,t_3,t_5,\ldots)$ of Eqs. (7.15)
as an implicit function:
$$
    x + \sum_{n=1}^\infty (2n+1) r_{n-1}(p) t_{2n+1}
    = \phi(p),                                           \tag 7.16
$$
where $\phi$ is an arbitrary function of one variable. Hence solving
this (transcendental) equation, one obtains a solution that depends
on an arbitrary function. Note that if $\phi$ has a Taylor expansion
at the origin, this arbitrary data can be absorbed into shift of the
time variables.

\subsection{7.3. Special solution for general $N$}
We construct a special solution for a general value of $N$ by the
method of Section 6. Let the $(f,g)$-pair be given by
$$
    f(\lambda,x) = \lambda^N/N, \quad
    g(\lambda,x) = x\lambda^{1-N}.                      \tag 7.17
$$
It is convenient to use
$$
    \P \=def \L^N/N, \quad
    \Q \=def \L^{1-N}\M                                    \tag 7.18
$$
rather than $\L$ and $\M$. This is essentially a change of canonical
variables:
$$
    \omega = d\L \wedge d\M = d\P \wedge d\Q.             \tag 7.19
$$
The Riemann-Hilbert problem for $\L$ and $\M$ is thus converted to
solving the equations
$$
    (\P)_{\le -1} = 0, \quad (\Q)_{\le -1} = 0            \tag 7.20
$$
for $\P$ and $\Q$. The first equation simply means that $\P$ is a
polynomial in $\lambda$:
$$
    \P = \lambda^N/N + p_2\lambda^{N-2} + \cdots + p_N.   \tag 7.21
$$
[We have slightly modified the parameterization in (7.2).] In view
of (7.18), $\Q$ is required to be a Laurent series of $\L$ of
the following form.
$$
    \Q = \sum_{n=1}^\infty nt_n \L^{n-N}
        +\sum_{i=0}^\infty v_{i+1}\L^{-i-N}               \tag 7.22
$$
(Actually, we shall see that the $v_1$-term disappears.) The second
equation of (7.20) then becomes
$$
    \sum_{n=1}^{N-1} nt_n \L^{n-N}
    +\sum_{i=0}^\infty v_{i+1} \L^{-i-N}
    +\sum_{k=N+1}^\infty kt_k (\L^{k-N})_{\le -1} = 0.
                                                          \tag 7.23
$$
Now multiply this equation with $\L^{N-n-1}d_\lambda\L$ or
$\L^{N+i-1}d_\lambda\L$ and take the residue.  In the first
and second sums of (7.23), only a single term survives after
this manipulation (recall Lemma B of Section 4). In the third
sum, the $k=N$ term disappears because $(\L^{k-N})_{\le -1} = 0$
for $k=N$, but the other terms give nontrivial constribution
in general.  Thus we have
$$
    nt_n + \sum_{k=N+1}^\infty k t_k
       \res[ (\L^{k-N})_{\le -1} \L^{N-n-1} d_\lambda \L ] =0
                                                             \tag 7.24
$$
for $n=1,\ldots,N-1$, and
$$
    v_{i+1} + \sum_{k=N+1}^\infty kt_k
       \res[ (\L^{k-N})_{\le -1} \L^{N+i-1} d_\lambda \L ] = 0
                                                             \tag 7.25
$$
for $i=0,1,\ldots$. We now argue that
\item{$\bullet$}
    Eqs. (7.24) may be thought of as a generalized hodograph
    transformation that determines the coefficients of $\P$ as
    implicit functions,
\item{$\bullet$}
    Eqs. (7.25) give the Laurent coefficients $v_{i+1}$ of $\Q$, and
\item{$\bullet$}
    Krichever's formula for the tau function and ``dispersionless
    analogues of Virasoro constraints" [3] can be reproduced in the
    present setting.

\subsection{7.4. Structure of Eqs. (7.24) and (7.25)}
We first consider Eqs. (7.24).  A basic tool is the residue
identities
$$
\align
    \res[ (\L^n)_{\le -1} \L^{m-1}d_\lambda\L]
  &= \frac{1}{m}\res[ (\L^n)_{\le -1} d_\lambda(\L^m)]              \\
  &= \frac{1}{m}\res[ (\L^n)_{\le -1} d_\lambda \B_m ]              \\
  &= \frac{1}{m}\res[ \L^n d_\lambda \B_m ]
                                                          \tag 7.26 \\
\endalign
$$
that we have used in Sections 4 and 5 [see (4.19) and (5.4)]. Note
that these residues give polynomials of $p_2$, $\ldots$, $p_N$.
In particular [see (5.7)],
$$
\align
    \res[ (\L)_{\le -1} \L^{N-n-1}d_\lambda \L ]
  &= \frac{1}{N-n} \res[ (\L^{N-n})_{\le -1} d_\lambda\L]           \\
  &= -q_{N-n+1}                                           \tag 7.27 \\
\endalign
$$

The last identity gives us a key to understand the role of
flat coordinates [25] in topological minimal models [4][5].
To see this, restrict part of time variables to special values:
$$
    t_{N+1} = \frac{1}{N+1}, \quad
    t_{N+2} = t_{N+3} = \cdots = 0.                      \tag 7.28
$$
Eqs. (7.24) then takes a very simple form:
$$
    nt_n - q_{N-n+1} = 0  \quad (n=1,\ldots,N-1).        \tag 7.29
$$
These are the fundamental relation in topological minimal models
that links time variables (which are identified with ``coupling
constants" [4][5]) of flow equations with flat coordinates.
The same relation can also be found in Krichever's interpretation.

Let us continue the analysis of Eqs. (7.24). To see how the unknown
functions $p_2$, $\ldots$, $p_N$ are to be determined, we have to know
the structure of the $q_n$'s as polynomials of the $p_n$'s. The theory
of flat coordinates [25] provides very detailed information on this
issue. For the moment, however, the following is sufficient.

\proclaim{Lemma C}
The invertible polynomial relation connecting $p_2$, $\ldots$, $p_N$
and $q_2$, $\ldots$, $q_N$ can be written
$$
    q_n = -p_n + (\text{\rm polynomials of } p_2,\ldots,p_N)  \tag 7.30
$$
for $n=2$, $\ldots$, $N$.
\endproclaim

\demo{Proof} The $i$-th power of $\L$, in general, can be expanded
in powers of $\lambda$ as:
$$
\align
    \L^i =& \lambda^i + iu_2\lambda^{i-2} + iu_3\lambda^{i-3} + \cdots \\
          & +\bigl( iu_j + (\text{polynomial of }u_2,\ldots,u_{j-1})
             \bigr) \lambda^{i-j}
            + \cdots,                                                  \\
\endalign
$$
therefore, letting $i=N$ and $j=n$, we have
$$
    Np_{n+1} = Nu_{n+1} + \bigl(\text{polynomial of }u_2,\ldots,u_{n}\bigr).
$$
Meanwhile, by the same expansion with $i=n$ and $j=n+1$,
$$
\align
    q_{n+1} &= -\frac{1}{n}\res \L^n d\lambda                     \\
            &= -u_{n+1} + (\text{polynomial of }u_2,\ldots,u_n).
\endalign
$$
{}From these relations, one can deduce (7.30). \qed
\enddemo

\noindent
Consequently, Eqs. (7.24) takes such a form as:
$$
\align
  & t_1 - (N+1)t_{N+1}q_N     + \sum_{k=N+2}^\infty kt_k\res[\cdots] = 0,
                                                                         \\
  & 2t_2 - (N+1)t_{N+1}q_{N-1}
                              + \sum_{k=N+2}^\infty kt_k\res[\cdots] = 0,
                                                                         \\
  & \cdots\cdots\cdots                                                   \\
  & (N-1)t_{N-1} - (N+1)t_{N+1}q_2
                              + \sum_{k=N+2}^\infty kt_k\res[\cdots] = 0,
                                                            \tag 7.31    \\
\endalign
$$
and the residues $\res[\cdots]$ are polynomials of $p_2$, $\ldots$, $p_N$
(or, equivalently, of $q_2$, $\ldots$, $q_N$). If $t_{N+1} \not= 0$,
one can rewrite these equations as:
$$
\align
  & q_N    = \frac{t_1}{(N+1)t_{N+1}}
            +\sum_{k=N+2}^\infty  \frac{kt_k \res[\cdots]}{(N+1)t_{N+1}}, \\
  & q_{N-1}= \frac{2t_2}{(N+1)t_{N+1}}
            +\sum_{k=N+2}^\infty  \frac{kt_k \res[\cdots]}{(N+1)t_{N+1}}, \\
  & \cdots\cdots\cdots                                                    \\
  & q_2    = \frac{(N-1)t_{N-1}}{(N+1)t_{N+1}}
            +\sum_{k=N+2}^\infty  \frac{kt_k \res[\cdots]}{(N+1)t_{N+1}}. \\
\endalign
$$
One can now use an ordinary implicit function theorem to ensure the
existence of functions $q_2$, $\ldots$, $q_N$ (hence $p_2$, $\ldots$,
$p_N$) that satisfy these equations. Note that these functions
depend only on the ratios $t_n/t_{N+1}$ ($n \not= N+1$) rather than
$t_n$ themselves. Thus:

\proclaim{Proposition 8}
Eqs. (7.24) has a solution that consists of homogeneous functions
$p_2$, $\ldots$, $p_N$ of degree zero. These functions are defined
in a domain where  $t_{N+1} \not= 0$ and $t_n/t_{N+1}$ ($n \not= N+1$)
are small.
\endproclaim

\noindent
Having such a solution of Eqs. (7.24), one can readily solve Eqs. (7.25) as:
$$
    v_{i+1} = -\sum_{k=N+1}^\infty kt_k
                \res[(\L^{k-N})_{\le -1} \L^{N+i-1}d_\lambda\L].
                                                           \tag 7.32
$$
The residues on the right hand side are polynomial functions of
$p_2$, $\ldots$, $p_N$, hence homogenous functions of degree zero.
Therefore:

\proclaim{Proposition 9}
A set of functions $v_{i+1}$ defined by (7.32) are homogeneous functions
of degree one and give a solution of Eqs. (7.25).  Further, $v_1 = 0$.
\endproclaim

\demo{Proof of last part} From (9.32) [and recalling (7.26)], we have
$$
\align
    v_1 &= -\sum_{k=N+1}^\infty kt_k
                \res[ (\L^{k-N})_{\le -1} \L^{N-1}d_\lambda\L]     \\
        &= -\sum_{k=N+1}^\infty kt_k
                \res[ (\L^N)_{\le -1} d_\lambda \B_{k-N}].         \\
\endalign
$$
The $N$-th power $\L^N$, however, is a polynomial of $\lambda$
because of the construction [see (7.21)]. Therefore $(\L^N)_{\le -1}=0$,
and the last residues should vanish.  This means $v_1=0$. \qed
\enddemo

\subsection{7.5. Tau function and nonlinear constraints}
Recall that the tau function is defined by
$$
    \frac{\del\log\tau}{\del t_n} = v_{n+1} \quad (n=1,2,\ldots).
                                                            \tag 7.33
$$
Since the right hand side of these equations are homogeneous
functions of degree one (Proposition 9), $\log\tau$ becomes
a homogeneous function of degree two (plus an integration constant).
Consequently,
$$
    \sum_{n=1}^\infty t_n \frac{\del\log\tau}{\del t_n}
    = 2\log\tau.
                                                            \tag 7.34
$$
Applying the operator $\sum_{n=1}^\infty t_n \del/\del t_n$ once again,
we have
$$
    \sum_{n,m=1}^\infty
        t_m t_n \frac{\del^2\log\tau}{\del t_m \del t_n}
    = 2\log\tau.                                            \tag 7.35
$$
These relations show an explicit form of the tau function:

\proclaim{Proposition 10}
The tau function, up to an integration constant, is given by
$$
    \log\tau = \frac{1}{2}\sum_{n=1}^\infty t_nv_{n+1} + \text{\rm const.}
                                                            \tag 7.36
$$
or, equivalently, by
$$
    \log\tau = \frac{1}{2}\sum_{n,m=1}^\infty
               t_m t_n \frac{\del v_{n+1}}{\del t_m} + \text{\rm const.}
                                                            \tag 7.37
$$
\endproclaim

The last formula, (7.37), has another expression. To see this,
recall that (Section 4)
$$
    \B_m = \L^m - \sum_{n=1}^\infty \frac{1}{n}
             \frac{\del v_{n+1}}{\del t_m} \L^{-n}.   \tag 7.38
$$
By virtue of this identity, (7.37) can be rewritten
$$
    \log\tau = \frac{1}{2}\res[
                  \left( \sum_{n=1}^\infty t_n \L^n \right)
                  d_\lambda
                  \left( \sum_{n=1}^\infty t_m \B_m \right)
               ].
                                                            \tag 7.39
$$
This exactly reproduces Krichever's formula of the tau function [3].

Krichever's dispersionless analogues of Virasoro constraints [3], too,
can be deduced as follows.

\proclaim{Proposition 11}
The tau function satisfies the constraints
$$
\align
  & \sum_{k=N+1}^\infty kt_k \frac{\del\log\tau}{\del t_{k-N}}
    +\frac{1}{2} \sum_{i=1}^{N-1} i(N-i) t_i t_{N-i} = 0,
                                                               \tag 7.40 \\
  & \sum_{k=1}^\infty kt_k \frac{\del\log\tau}{\del t_k} = 0,
                                                               \tag 7.41 \\
  & \sum_{k=1}^\infty kt_k \frac{\del\log\tau}{\del t_{k+mN}}
    +\frac{1}{2} \sum_{i=1}^{mN-1}
        \frac{\del\log\tau}{\del t_i} \frac{\del\log\tau}{\del t_{mN-i}}
                                                               \tag 7.42 \\
\endalign
$$
where $m$ ranges over $m=1,2,\ldots$.
\endproclaim

\demo{Proof} We first derive (7.40) from Eqs. (7.24) and (7.25).
Recall, again, the identity (see Section 5)
$$
    \res[\L^n d_\lambda \B_m]
    = \frac{\del v_{n+1}}{\del t_m} = \frac{\del v_{m+1}}{\del t_n}
    = \frac{\del^2 \log\tau}{\del t_m \del t_n}.            \tag 7.43
$$
Because of these identities, Eqs. (7.24) and (7.25) become differential
equations for $\log\tau$:
$$
\align
  & (N-n)nt_n + \sum_{k=N+1}^\infty kt_k
       \frac{\del^2 \log\tau}{\del t_{N-n} \del t_{k-N}} = 0,  \tag 7.44 \\
  & (N+i)\frac{\del\log\tau}{\del t_i}
    +\sum_{k=N+1}^\infty kt_k
       \frac{\del^2 \log\tau}{\del t_{N+i} \del t_{k-N}} = 0,  \tag 7.45 \\
\endalign
$$
where the indices $n$ and $i$ range over $n=1,\ldots,N-1$ and
$i=1,2,\ldots$. Eq. (7.45), after replacing $i \to i-N$, can be rewritten
$$
    \frac{\del}{\del t_i}\left(
       \sum_{k=N+1}^\infty kt_k \frac{\del\log\tau}{\del t_{k-N}} \right)
    = 0,                                                       \tag 7.46
$$
which means that $\sum_{k=N+1}^\infty kt_k \del\log\tau/\del t_{k-N}$
does not depend on $t_{N+1}$, $t_{N+2}$, $\ldots$. Eqs. (7.44) can be
rewritten, by similar calculations, as
$$
    \frac{\del}{\del t_{N-n}} \left(
      \sum_{k=N+1}^\infty kt_k \frac{\del\log\tau}{\del t_{k-N}}
      +\frac{1}{2}\sum_{i=1}^{N-1} i(N-i) t_i t_{N-i}
    \right) = 0.                                               \tag 7.47
$$
These relations show that
$$
    \sum_{k=N+1}^\infty kt_k \frac{\del\log\tau}{\del t_{k-N}}
    +\frac{1}{2}\sum_{i=1}^{N-1} i(N-i) t_i t_{N-i}
    = \text{const.}
$$
The left hand side of this equality, however, is a homogeneous function
of degree two, hence the constant on the right hand side has to vanish.
One can thus derive (7.40). To derive (7.41) and (7.42), note that
Eqs. (7.24) and (7.25) are simply a restatement of the second equation
of (7.20).  Actually, $\P$ and $\Q$ obeying (7.20) also satisfy an
infinite number of similar equations such as
$$
    ( \P^{m+1} \Q )_{\le -1} = 0  \quad (m=0,1,2,\ldots),     \tag 7.48
$$
and each of these equations give rise to a set of relations like
Eqs. (7.24) and (7.25).  Starting from these relations, one can
deduce (7.41) and (7.42) in the same way as we have derived
(7.40).  This completes the proof. \qed
\enddemo

The above proof shows that there are actually more constraints
that our tau function are satisfying.  Besides (7.48), in fact,
we have
$$
    ( \P^{m+1} \Q^n )_{\le -1} = 0 \quad (m \ge -1, n \ge 2),
                                                              \tag 7.49
$$
and accordingly a corresponding set of nonlinear constraints. This
observation reminds us of the ``W constraints" in the $d<1$ string
theory [34][35] and ``twisted $W_\infty$ constraints" in a $d=1$
version [19]. We shall show a more systematic interpretation of
these constraints in Section 8, exploiting SDiff(2) ($= w_\infty$)
symmetries of our hierarchy.

\subsection{7.6. Deformations}
The previous solution has a family of deformations
generated by the $(f,g)$-pair
$$
    f(\lambda,x) = \lambda^N/N, \quad
    g(\lambda,x) = \lambda^{1-N}x + h(\lambda),          \tag 7.50
$$
where $h(\lambda)$ is an arbitrary function with Laurent expansion
$$
    h(\lambda) = \sum_{k=-\infty}^\infty h_k \lambda^k.
                                                         \tag 7.51
$$
The $(\P,\Q)$-pair is now given by
$$
    \P = \L^N/N, \quad \Q = \L^{1-N}\M + h(\L).          \tag 7.52
$$
Eqs. (7.24) and (7.25) are also deformed:
$$
\align
  & nt_n + h_{n-N} + \sum_{k=N+1}^\infty (kt_k+h_{k-N})
          \res[ (\L^{k-N})_{\le -1} \L^{N-n-1} d_\lambda \L ] = 0,
                                                             \tag 7.53 \\
  & v_{i+1} + h_{-i-N} + \sum_{k=N+1}^\infty (kt_k+h_{k-N})
          \res[ (\L^{k-N})_{\le -1} \L^{N+i-1} d_\lambda \L ] = 0.
                                                             \tag 7.54 \\
\endalign
$$
Previous calculations can be mostly carried over to this case,
except that the $v_1$-term now does not vanish in general,
$$
    v_1 = - h_{-N}.                                       \tag 7.55
$$
(This is, however, a redundant degree of freedom as we have mentioned in
Section 4.) If $N=2$, Eqs. (7.53) reduces to a single equation:
$$
    x + h_{-1} + \sum_{k=N+1}^\infty (kt_k + h_{k-2})
              \res[(\L^{k-1})_{\le -1} d_\lambda\L] = 0.  \tag 7.56
$$
The residues in the last part can be written
$$
\align
    \res[ (\L^{k-1})_{\le -1} d_\lambda \L ]
    &= \res[ \L^{k-2} d\lambda]                                     \\
    &= \cases
         r_{n-1}(p)   &\text{if $k=2n+1$},  \\
         0            &\text{if $k=2n$}.    \\
       \endcases                                          \tag 7.57 \\
\endalign
$$
Thus Eq. (7.56) is essentially the same ``hodograph" relation as
Eq. (7.16). The arbitrary function $\phi(p)$ is connected with
the Riemann-Hilbert data $h(\lambda)$ as:
$$
    \phi(p) = -\res h(\L)d\lambda.                        \tag 7.58
$$

The tau function of the deformed solutions can be readily determined,
because the presence of $h(\lambda)$ simply affects as shifting $t_n$'s
and $v_{i+1}$'s by constants:

\proclaim{Proposition 12}
Let $\tau_0=\tau_0(t)$ be the tau function of the undeformed solution
($h=0$). The tau function $\tau_h=\tau_h(t)$ of the deformed solution
is then given by
$$
\align
    \tau_h(t) =& \text{\rm const.}
                 \exp\left( -\sum_{n=1}^\infty h_{-n-N}t_n \right)    \\
               &\times \tau_0(t_1 + h_{1-N}, t_2 + \frac{h_{2-N}}{2},
                              \ldots, t_k + \frac{h_{k-N}}{k}, \ldots).
                                                           \tag 7.59  \\
\endalign
$$
\endproclaim

In the context of topological minimal models, insertion of
$h(\lambda)$ amounts to ``perturbations" of the model [4][5].
Of course, as we have seen above, these deformations are actually
absorbed into the time variables $t_n$ of the hierarchy. In the
same context, Eqs. (7.24) and (7.53) may be interpreted as
``Landau-Ginzburg equations," i.e., ``topological" analogues
of $d<1$ string equations [36]. Unlike the latter, these
topological analogues contain no derivatives of unknown functions.
One can therefore resort to the implicit function theorem to
ensure the existence of solutions.

\section{8. SDiff(2) symmetries}

\subsection{8.1. Symmetries in terms of $(\L,\M)$}
The method of construction of symmetries is based upon the same
principle as the case of the self-dual vacuum Einstein equations
[37][38]. We have established a correspondence between
$(\L,\M)$ and $(f,g)$. The SDiff(2) group structure in the data
$(f,g)$ gives rise to transformations of a solution to another by,
say, the right action of a given SDiff(2) group element. To find
their infinitesimal form, we consider a one-parameter family of
transformations $(\L,\M)\to(\L(\epsilon),\M(\epsilon))$ generated by
the right action
$$
    (f,g) \longrightarrow
    (f,g) \circ \exp(-\epsilon \{F, \cdot\}),               \tag 8.1
$$
where $\{F,\cdot\}$ is a Hamiltonian vector field,
$$
    \{F,\cdot\} \=def \frac{\del F}{\del\lambda}\frac{\del}{\del x}
                     -\frac{\del F}{\del x}\frac{\del}{\del\lambda},
                                                            \tag 8.2
$$
and $F(\lambda,x)$ is assumed to have the same analyticity properties
as $f(\lambda,x)$ and $g(\lambda,x)$. We then calculate the transformed
pair $(\L(\epsilon),\M(\epsilon))$ to the first order of $\epsilon$:
$$
\align
    \L(\epsilon) &= \L + \epsilon \delta \L + O(\lambda^2),          \\
    \M(\epsilon) &= \M + \epsilon \delta \M + O(\lambda^2).
                                                            \tag 8.3 \\
\endalign
$$
The coefficients $\delta \L$ and $\delta \M$ define a linear operator
$\delta=\delta_F$ that represents an infinitesimal symmetry of
the SDiff(2) KP hierarchy. By definition, $\delta_F$ acts on any
function of $\L$ and $\M$ as an abstract derivation,
$$
    \delta_F G(\L,\M) = \frac{\del G}{\del\L}\delta_F \L
                       +\frac{\del G}{\del\M}\delta_F \M    \tag 8.4
$$
whereas leaves invariant the independent variables of the hierarchy,
$$
    \delta_F t_n = \delta_F x = \delta_F \lambda = 0.       \tag 8.5
$$
(This is a formal way to understand infinitesimal symmetries of
differential equations in the language of differential algebras
[39].)

For the self-dual vacuum Einstein equation, infinitesimal symmetries
thus constructed have an explicit and compact expression [38];
this is also the case for the present situation.  Since precise
calculations are only required to the first order of $\epsilon$,
one can write the Riemann-Hilbert problem for $\L(\epsilon)$ and
$\M(\epsilon)$ as:
$$
\align
    f\left(
      \lambda + \epsilon F_x(\lambda,x) + O(\epsilon^2),
      x -\epsilon F_\lambda(\lambda,x)  + O(\epsilon^2)
    \right)|_{\lambda\to\L(\epsilon), x\to\M(\epsilon)} &           \\
          = \Lhat + \epsilon\delta\Lhat + O(\lambda^2), &
                                                                    \\
    g\left(
      \lambda + \epsilon F_x(\lambda,x) + O(\epsilon^2),
      x -\epsilon F_\lambda(\lambda,x)  + O(\epsilon^2)
    \right)|_{\lambda\to\L(\epsilon), x\to\M(\epsilon)} &           \\
          = \Mhat + \epsilon\delta\Mhat + O(\lambda^2), &
                                                           \tag 8.6 \\
\endalign
$$
where $F_\lambda(\lambda,x) \=def \del F(\lambda,x) / \del\lambda$, etc.
as usual; note that the right hand side are also affected by the change
of the data $(f,g)$. From the $\epsilon$-terms, one obtains a set of
equations that should determine $\delta\L$, $\delta\M$, $\delta\Lhat$
and $\delta\Mhat$. In a matrix form, these equations can be written
$$
    \left(\matrix
      \dfrac{\del f(\L,\M)}{\del\L} & \dfrac{\del f(\L,\M)}{\del\M} \\
      \dfrac{\del f(\L,\M)}{\del\L} & \dfrac{\del f(\L,\M)}{\del\M} \\
    \endmatrix\right)
    \left(\matrix
      \delta\L + \dfrac{\del F(\L,\M)}{\del\M} \\
      \delta\M - \dfrac{\del F(\L,\M)}{\del\L} \\
    \endmatrix\right)
  = \left(\matrix
      \delta\Lhat \\
      \delta\Mhat \\
    \endmatrix\right).                                   \tag 8.7
$$
Now the situation is very similar to the proof of Proposition 7,
except that $\del\L/\del t_n$ etc. therein are replaced as:
$$
\align
    \frac{\del\L}{\del t_n} \longrightarrow
        \delta\L + \frac{\del F(\L,\M)}{\del\M},               \quad &
    \frac{\del\Lhat}{\del t_n} \longrightarrow \delta\Lhat,
                                                                    \\
    \frac{\del\M}{\del t_n} \longrightarrow
        \delta\M - \frac{\del F(\L,\M)}{\del\L},               \quad &
    \frac{\del\Mhat}{\del t_n} \longrightarrow \delta\Mhat.
                                                          \tag 8.8  \\
\endalign
$$
After eliminating the derivative matrix of $(f,g)$ by $(\L,\M)$, one has
$$
\align
    \frac{\del\M}{\del x}
      \left(\delta\L + \frac{\del F(\L,\M)}{\del\M}\right)
   -\frac{\del\L}{\del x}
      \left(\delta\M - \frac{\del F(\L,\M)}{\del\L}\right)
 &= \frac{\del\Mhat}{\del x}\delta\Lhat
   -\frac{\del\Lhat}{\del x}\delta\Mhat,
                                                                     \\
    \frac{\del\M}{\del\lambda}
      \left(\delta\L + \frac{\del F(\L,\M)}{\del\M}\right)
   -\frac{\del\L}{\del\lambda}
      \left(\delta\M - \frac{\del F(\L,\M)}{\del\L}\right)
 &= \frac{\del\Mhat}{\del\lambda}\delta\Lhat
   -\frac{\del\Lhat}{\del\lambda}\delta\Mhat,            \tag 8.9  \\
\endalign
$$
or, in a more compact form,
$$
\align
    \frac{\del\M}{\del x}\delta\L
   -\frac{\del\L}{\del x}\delta\M
   -\frac{\del}{\del x}F(\L,\M)
 &= \frac{\del\Mhat}{\del x}\delta\Lhat
   -\frac{\del\Lhat}{\del x}\delta\Mhat,
                                                                  \\
    \frac{\del\M}{\del\lambda}\delta\L
   -\frac{\del\L}{\del\lambda}\delta\M
   -\frac{\del}{\del\lambda}F(\L,\M)
 &= \frac{\del\Mhat}{\del\lambda}\delta\Lhat
   -\frac{\del\Lhat}{\del\lambda}\delta\Mhat.            \tag 8.10
                                                                  \\
\endalign
$$
The $(\quad)_{\le -1}$-part of the last equations give
$$
\align
  & \frac{\del\M}{\del x}\delta\L
    -\frac{\del\L}{\del x}\delta\M
    -\frac{\del}{\del x}F(\L,\M)_{\le -1} = 0,
                                                                   \\
  & \frac{\del\M}{\del\lambda}\delta\L
    -\frac{\del\L}{\del\lambda}\delta\M
    -\frac{\del}{\del\lambda}F(\L,\M)_{\le -1} =0.       \tag 8.11 \\
\endalign
$$
which can easily be solved with respect to $\delta\L$ and $\delta\M$
because the coefficient matrix is unimodular. One thus arrives at
the following result.

\proclaim{Proposition 13} The infinitesimal symmetries $\delta_F\L$ and
$\delta_F\M$ are given by
$$
\align
    & \delta_F\L = \{ F(\L,\M)_{\le -1}, \L \},                    \\
    & \delta_F\M = \{ F(\L,\M)_{\le -1}, \M \}.          \tag 8.12 \\
\endalign
$$
\endproclaim

\subsection{8.2. Symmetries in terms of $v_{i+1}$}
Symmetries of the self-dual vacuum Einstein equation are further
extended to potentials called Plebanski's ``key functions" [38].
For several reasons, it is $v_2$ that should correspond to the key
functions.  For comparison, we now show the action of $\delta_F$ on
the Laurent coefficients $v_i$.  The formula for $\delta_F v_2$
is indeed reminiscent of a similar formula for the key functions.

\proclaim{Proposition 14} For $i=1,2,\ldots$, $\delta_F v_{i+1}$ are
given by
$$
    \delta_F v_{i+1} = -\res F(\L,\M)d_\lambda B_i.      \tag 8.13
$$
In particular,
$$
    \delta_F v_2 = -\res F(\L,\M)d\lambda.               \tag 8.14
$$
\endproclaim

\rmproclaim{Remark} The above formula is actually valid for $i=0$
and gives
$$
    \delta_F v_1 = 0                                    \tag 8.15
$$
if one retains the $v_1$-term in the definition of the hierarchy
(Section 4).  This means that the above infinitesimal symmetries
leave invariant $v_1$. Nevertheless we have seen in the previous
section that a nonzero $v_1$-term can be generated in the course
of solving a Riemann-Hilbert problem. This apparent discrepancy
is due to the special form of the one-parameter family of deformations,
(8.1), for which we have assumed the existence of a single-valued
generating function $F(\lambda,x)$.
\endrmproclaim

\demo{Proof of Proposition 14} The essence is the same as the proof of
Proposition 4. Since $\delta_F$ is a derivation like $\del/\del t_n$,
the chain rule applies:
$$
    \delta_F \M
    = \left( \frac{\del\M}{\del\L} \right)_{t,v\;\text{\rm fixed}}
        \delta_F \L
     +\sum_{i=1}^\infty (\delta_F v_{i+1})\L^{-i},         \tag 8.16
$$
and from the $\L^{-i}$-term, one has
$$
\align
  \delta_F v_{i+1}
  &= \res \L^i
       \left[
         \delta_F\M
        -\left( \frac{\del\M}{\del\L} \right)
                  _{t,v\;\text{\rm fixed}}\delta_F\L
       \right] d_\lambda\L
                                                                        \\
  &= \res \L^i [ \delta_F\M d_\lambda\L - \delta_F\L d_\lambda\M ]
                                                                        \\
  &= \res \L^i [ \{ F(\L,\M)_{\le -1},\M \}d_\lambda\L
                -\{ F(\L,\M)_{\le -1},\L \}d_\lambda\M ]
                                                                        \\
  &= \res \L^i d_\lambda F(\L,\M)_{\le -1},
                                                          \tag 8.17     \\
\endalign
$$
and finally, due to the properties of formal residues (Lemmas A and B),
$$
\align
  &= \res \B_i d_\lambda F(\L,\M)                                       \\
  &= -\res F(\L,\M) d_\lambda\B_i.                                      \\
\endalign
$$
This proves (8.13).
\qed\enddemo

\subsection{8.3. Symmetries extended to tau function}
Since our definition of the tau function is always accompanied
with an integration constant, symmetries at the level of
$\L$ and $\M$ do not automatically extend to the tau function.
Besides, if such an extension exists,
it is not ensured whether the extension has a simple form.
In fact, the following result shows that such an extension does
exist with a very simple expression.

\proclaim{Proposition 10} The infinitesimal symmetries $\delta_F$
of the $(\L,\M)$-pair can be consistently extended to the tau
function by defining
$$
    \delta_F\log\tau = - \res F^x(\L,\M)d_\lambda\L,        \tag 8.18
$$
where $F^x(\lambda,x)$ is a primitive function of $F(\lambda,x)$
normalized as
$$
    F^x(\lambda,x) \=def \int_0^x F(\lambda,y)dy.           \tag 8.19
$$
``Consistency" means that the following relation is satisfied.
$$
    \frac{\del}{\del t_n}\delta_F\log\tau
    = \delta_F \frac{\del\log\tau}{\del t_n}.               \tag 8.20
$$
\endproclaim

\demo{Proof} The right hand side of the consistency relation has been
calculated (Proposition 13):
$$
    \delta_F\frac{\del\log\tau}{\del t_n}
    = \delta_F v_{n+1}
    = -\res F(\L,\M)d_\lambda\B_n.                          \tag 8.21
$$
To calculate the other side, we introduce a set of
functions $\F_i=\F_i(t,v)$ of $t$ and $v=(v_2,v_3,\ldots)$ by
the Laurent expansion
$$
    F^x(\L,\M)
    = F^x(\L, \sum_{n=1}^\infty nt_n\L^{n-1}
               + \sum_{i=1}^\infty v_{i+1}\L^{-i-1})
    = \sum_{i=-\infty}^\infty \F_i \L^i                     \tag 8.22
$$
with respect to $\L$. The $t$-dependence of $\F_i$ comes only from
the $t$'s and $v$'s included in $\M$ of $F^x(\L,\M)$; in this
definition, $\L$ simply plays the role of an independent parameter.
Therefore, to calculate the $t$-derivatives of $\F_i$, one may
temporally consider $\L$ as a constant, and differentiate both hand
sides of the above relation. Thus one has
$$
\align
    \sum_{i=-\infty}^\infty \frac{\del\F_i}{\del t_n} \L^i
  & = \frac{\del F^x(\L,\M)}{\del\M}
      \left( \frac{\del\M}{\del t_n} \right)
          _{\L,t_m(m \not= n)\;\text{\rm fixed}}               \\
  & = F(\L,\M) \left( \frac{\del\M}{\del t_n} \right)
                   _{\L,t_m(m \not= n)\;\text{\rm fixed}}.     \\
\endalign
$$
Now one can apply the method of proof of Proposition 4 once again.
First, the last line can be continued as:
$$
    = F(\L,\M)
     \left[
         \frac{\del\M}{\del t_n}
        -\left( \frac{\del\M}{\del\L} \right)_{t,v\;\text{\rm fixed}}
           \frac{\del\L}{\del t_n}
     \right].                                             \tag 8.23
$$
Then from the $\L^{-1}$-term,
$$
\align
    & \frac{\del}{\del t_n}\delta_F\log\tau
       = -\frac{\del \F_{-1}}{\del t_n}                         \\
    &= -\res F(\L,\M) \left[
             \frac{\del\M}{\del t_n}
            -\left( \frac{\del\M}{\del\L} \right)_{t,v\;\text{\rm fixed}}
               \frac{\del\L}{\del t_n}
        \right] d_\lambda\L                                     \\
\endalign
$$
and this can be calculated just as in the proof of Proposition 4:
$$
\align
    &= -\res F(\L,\M) \left[
            \frac{\del\M}{\del t_n} d_\lambda \L
           -\frac{\del\L}{\del t_n} d_\lambda \M
        \right]
                                                                     \\
    &= -\res F(\L,\M) d_\lambda \B_n.                     \tag 8.24  \\
\endalign
$$
Eqs. (8.21) and (8.24) give the identical results, and this is
exactly the consistency relation.
\qed
\enddemo

\subsection{8.4. Commutation relations of symmetries}
Our final task is to calculate the commutation relations of these
infinitesimal symmetries.

\proclaim{Proposition 16} For any two generating functions
$F_1=F_1(\lambda,x)$ and $F_2=F_2(\lambda,x)$, the infinitesimal
symmetries $\delta_{F_1}$ and $\delta_{F_2}$ obey the commutation
relations
$$
    \left[ \delta_{F_1},\delta_{F_2} \right]\log\tau
    = \delta_{\{F_1,F_2\}}\log\tau + c(F_1,F_2)          \tag 8.25
$$
for the tau function and
$$
    \left[ \delta_{F_1},\delta_{F_2} \right] \K
    = \delta_{\{F_1,F_2\}} \K                            \tag 8.26
$$
for $\K=\L,\M$, where
$$
    c(F_1,F_2) \=def \res F_1(\lambda,0)dF_2(\lambda,0).     \tag 8.27
$$
\endproclaim

\demo{Proof} (8.26) is an immediate consequence of (8.25) and the
consistency in the sense of Proposition 14.  We shall only prove
(8.25). Without loss of generality, we may assume that
$$
    F_1 = f_1(\lambda)x^j, \quad F_2 = f_2(\lambda)x^k   \tag 8.28
$$
where $j$ and $k$ are nonnegative integers. Accordingly,
$$
\align
  & F_1^x = \frac{f_1(\lambda)x^{j+1}}{j+1}, \quad
    F_2^x = \frac{f_2(\lambda)x^{k+1}}{k+1},                          \\
  & \{ F_1,F_2 \}^x
    = \cases
        \dfrac{(kf_1'f_2-jf_1f_2')x^{j+k}}{j+k} &\text{\rm if $j+k>0$},\\
        0                                      &\text{\rm if $j+k=0$}. \\
      \endcases
                                                         \tag 8.29    \\
\endalign
$$
Let us examine the action of the commutator.  From the construction,
$$
    \left[ \delta_{F_1},\delta_{F_2} \right]\log\tau
    = -\delta_{F_1} \res F_2^x(\L,\M) d_\lambda\L
      +\delta_{F_2} \res F_1^x(\L,\M) d_\lambda\L.       \tag 8.30
$$
Note that the situation is the same as the proof of the previous
proposition; one has to calculate a derivative of a formal residue.
The only difference is that we now have a more abstract derivation
$\delta_F$ rather than $\del/\del t_n$.  For the first term
on the right hand side, thus,
$$
    \delta_{F_1}\res F_2^x(\L,\M) d_\lambda\L
    = \res F_2(\L,\M)[ \delta_{F_1}\M d_\lambda\L
                      -\delta_{F_1}\L d_\lambda\M ]
$$
and by the method of proof of Proposition 4, again,
$$
\align
    &= \res F_2(\L,\M) d_\lambda F_1(\L,\M)_{\le -1}              \\
    &= \res F_2(\L,\M)_{\ge 0} d_\lambda F_1(\L,\M)_{\le -1}.
                                                        \tag 8.31 \\
\endalign
$$
Similarly,
$$
    \delta_{F_2}\res F_1^x(\L,\M) d_\lambda\L
    = \res F_1(\L,\M)_{\ge 0} d_\lambda F_2(\L,\M)_{\le -1}.
                                                        \tag 8.32
$$
{}From (8.30)-(8.32),
$$
\align
  & \left[ \delta_{F_1},\delta_{F_2} \right]\log\tau              \\
  &= \res\left[
        -F_2(\L,\M)_{\ge 0} d_\lambda F_1(\L,\M)_{\le -1}
        +F_1(\L,\M)_{\ge 0} d_\lambda F_2(\L,\M)_{\le -1}
     \right]
                                                                  \\
  &= \res\left[
         F_1(\L\,\M)_{\le -1} d_\lambda F_2(\L,\M)_{\ge 0}
        +F_1(\L,\M)_{\ge 0}   d_\lambda F_2(\L,\M)_{\le -1}
     \right]
                                                                  \\
  &= \res F_1(\L,\M) d_\lambda F_2(\L,\M).              \tag 8.33 \\
\endalign
$$
Now suppose that $j+k=0$ (i.e., $j=k=0$). Then from Eq. (8.33),
$$
\align
    & \left[ \delta_{F_1},\delta_{F_2} \right] \log\tau
      = \res f_1(\L) d_\lambda f_2(\L)                             \\
    & = \res f_1(\lambda) d f_2(\lambda)
      = c(F_1,F_2)                                       \tag 8.34 \\
\endalign
$$
whereas
$$
    \delta_{\{F_1,F_2\}}\log\tau = 0,                      \tag 8.35
$$
hence (8.25) is satisfied for this case.
Meanwhile, if $j+k>0$,
$$
\align
    & \left[ \delta_{F_1},\delta_{F_2} \right]\log\tau
      = \res f_1(\L)\M^j
        [ f_2'(\L)\M^k d_\lambda\L +k f_2(\L) \M^{k-1}d_\lambda\M]
                                                                    \\
    & = \res\left[
          f_1(\L)f_2'(\L)\M^{j+k}d_\lambda\L
          +\frac{k}{j+k}f_1(\L)f_2(\L)d_\lambda( \M^{j+k} )
        \right]
                                                                    \\
    & = \res\left[
          f_1(\L)f_2'(\L)\M^{j+k}d_\lambda\L
          -\frac{k}{j+k}\M^{j+k}d_\lambda\bigl( f_1(\L)f_2(\L) \bigr)
        \right]
                                                                    \\
    &= -\res\left[
           \frac{k}{j+k}f_1'(\L)f_2(\L)
          -\frac{j}{j+k}f_1(\L)f_2'(\L)
        \right] \M^{j+k} d_\lambda \L
                                                                    \\
    & = \delta_{\{F_1,F_2\}}\log\tau.                  \tag 8.36    \\
\endalign
$$
Since $c(F_1,F_2) = 0$, this shows that (8.25) is satisfied for
this case as well.  This completes the proof.
\qed
\enddemo

We have thus observed that the infinitesimal symmetries at the
level of the tau function exhibit anomalous commutation relations.
The anomalous term, $c(F_1,F_2)$, is a cocycle of the SDiff(2)
algebra on a cylinder, hence gives rise to a central extension.
This result, too, advocates that our definition of the tau function
is an appropriate one as an analogue of the tau function of the
ordinary KP hierarchy.  One should  note that anomalous
commutation relations of the SDiff(2) version are limited to the
``spin-1" sector (i.e., $\delta_F$ with $F=F(\lambda)$) of the
SDiff(2) algebra. This is in contrast with the case of the ordinary
KP hierarchy; anomalous commutation relations therein take place in
all sector of the $gl(\infty)$ algebra [2] or of the
$\hat{W}_\infty$ algebra [19].

Cocycles of SDiff(2) algebras on various surfaces are classified
by physicists [40][41][42][43].
According to their analysis, there are $2g$ linearly independent
cocycles on a genus $g$ surface. Since a cylinder $S^1 \times \bR^1$
may be thought of as a genus $g=1/2$ surface, the space of nontrivial
cocycle should be one-dimensional. Our cocycle gives a realization of
such a cocycle.

\subsection{8.5. SDiff(2) constraints of topological minimal models}
We have seen in Section 7 that the tau function $\tau=\tau_0(t)$
of the undeformed ($h=0$) solution constructed therein satisfies
an infinite set of nonlinear constraints.  Actually, these constraints
have a very simple interpretation in terms of the SDiff(2) symmetries
as follows. Let us start from the basic relations
$$
    \left( \L^{(m-1)N+n(1-N)} \M^n \right)_{\le -1}
    = \text{const.}\left( \P^{m-1} \Q^{n} \right)_{\le -1} = 0
                                                        \tag 8.37
$$
for $m \ge -1$ and $n \ge 0$. Each of these relations is equivalent
to the requirement that the equations
$$
    \res[ \L^{(m-1)N+n(1-N)} \M^n d_\lambda \B_i] = 0
                                                        \tag 8.38
$$
be satisfied for $i \ge 1$, because the $\B_i$'s ($i \ge 1$) form a
basis of the vector subspace of polynomials of $\lambda$ in the space
of Laurent series. By Proposition 14, one can rewrite Eq. (8.38)
in terms of SDiff(2) symmetries as:
$$
    \delta_{\lambda^{(m-1)N+n(1-N)}x^n}v_{i+1} = 0.      \tag 8.39
$$
Further, since $v_{i+1}=\del\log\tau/\del t_i$ and the infinitesimal
symmetries commute with $\del/\del t_i$ in the sense of Proposition 15,
Eq. (8.39) becomes
$$
    \frac{\del}{\del t_i}\delta_{\lambda^{(m-1)N+n(1-N)}x^n}\log\tau
    = 0.                                                 \tag 8.40
$$
Consequently,
$$
    \delta_{\lambda^{(m-1)N+n(1-N)}x^n}\log\tau = \text{const.}
                                                         \tag 8.41
$$
The left hand side of this equation can be written in a residue form
(Proposition 15) as:
$$
    \delta_{\lambda^{(m-1)N+n(1-N)}x^n}\log\tau
    = -\frac{1}{n+1}\res[ \L^{(m-1)N+n(1-N)} \M^{n+1} d_\lambda\L],
                                                         \tag 8.42
$$
and this is a homogeneous function of degree $n+1$ ($\not= 0$). Hence
the constant on the right hand side of (8.41) has to vanish. To summarize:

\proclaim{Proposition 17}  The tau function $\tau=\tau_0(t)$
of Proposition 10 satisfies the constraints
$$
    \delta_{\lambda^{(m-1)N+n(1-N)}x^n}\log\tau
    = -\frac{1}{n+1}\res[ \L^{(m-1)N+n(1-N)} \M^{n+1} d_\lambda\L]
    = 0                                                 \tag 8.43
$$
for $m \ge -1$ and $n \ge 0$.
\endproclaim

Eqs. (8.43) give a compact expression of the constraints mentioned
in Section 7. If $n=0$, one obtains the obvious relations
$$
    v_{(m-1)N} = \frac{\del\log\tau}{\del t_{(m-1)N}} = 0,
                                                        \tag 8.44
$$
which are satisfied by any solution of the $N$-reduced hierarchy.
If $n=1$,  Eqs. (8.42) are nothing but Krichever's dispersionless
analogue of Virasoro constraints. The others for $n \ge 2$ give
higher constraints equivalent to (7.49).

\section{9. Conclusion}

\noindent
Our SDiff(2) KP hierarchy is, in many aspects, very similar to the
ordinary KP hierarchy.  Their differences should be ultimately
due to the difference of the underlying Lie algebras, i.e., the SDiff(2)
algebra and the $\gl(\infty)$ algebra. The $S$ function
and the $\tau$ function both have counterparts in the KP hierarchy.
Infinitesimal symmetries are constructed and shown to exhibit
anomalous commutation relations at the level of the tau function.
These show a remarkable similarity between the two distinct hierarchies.

Besides the similarity with the KP hierarchy, the SDiff(2) KP hierarchy
shares a number of characteristics with the self-dual vacuum Einstein
equations and its 3-d reductions [6][7]. The Riemann-Hilbert problem in
the SDiff(2) group is a key to connect this hierarchy with the
minitwistor theory [13][14][15][16].

This double nature of the SDiff(2) KP hierarchy can also be seen
in its Toda version [17].  We therefore expect these SDiff(2)
hierarchies to play the role of a bridge that connects two distinct
families of nonlinear integrable systems, i.e., soliton equations
(most of which live in two dimensions) and self-duality equations
(which live in four dimensions). In this respect, an intriguing
problem will be to pursue Orlov's approach to the KP hierarchy
[18][19] as a ``noncommutative minitwistor theory." If this turns
out to be successful, the next step would be naturally to construct
a noncommutative analogue of full twistor theory in four dimensions
that should reproduce the twistor theory of the self-dual vacuum
Einstein equation [26] as a ``quasi-classical'' limit. We do not
know what a corresponding nonlinear ``integrable'' system looks like.

Our knowledge on special solutions of the SDiff(2) KP hierarchy is
very limited. As Krichever [3] pointed out in his framework of
dispersionless Lax equations, topological minimal models [4][5]
give such special solutions. We have presented a characterization
of these solutions in the language of the Riemann-Hilbert problem,
and found a set of nonlinear constraints satisfied by the corresponding
tau function.  These constraints include Krichever's ``dispersionless
analogues of Virasoro constraints" [3], hence may be called ``SDiff(2)
(or $w_\infty$) constraints."

An important problem still left open is to find a geometric structure
like the infinite dimensional Grassmannian manifold [1][2][20].  A useful
expression of the tau function will be obtained from such a geometric
structure. For the SDiff(2) Toda equation, Saveliev and his collaborators
[10] indeed presented an expression of solutions that seems to provide a
hint to this problem. This issue should also be related to some quantum
field theory like the free fermion theory emerging in the KP hierarchy [2].
In view of the relation to topological minimal models, an underlying
field theory will be a kind of topological field theory in a generalized
sense.

\section{Acknowledgements}

\noindent
We would like to express our gratitude to I. Bakas, J.D. Finley,
I.M. Krichever, Q-Han Park, A.G. Reiman and M.V. Saveliev for
discussions and a number of useful comments.  We are also grateful
to the staff of the Research Institute for Mathematical Sciences.

\section{Note added}

\noindent
After completing this work, we are informed of papers by
Kodama and Gibbons [44]. They deal with the same hierarchy (and
a Toda version) and present remarkable results on special
solutions.  We thank Takahiro Shiota for this information.

%%%% references %%%%%%%%%%%%%%%%%%%%%%%%%%%%%%%%%%%%%%%%%%%%%%%%%%%%%%%%%
\references

%%%  US-Japan seminar
\no{1}{Sato, M., and Sato, Y.,}
in \book{Nonlinear Partial Differential
Equations in Applied Sciences}, P.D. Lax, H. Fujita, and G. Strang (eds.)
(North-Holland, Amsterdam, and Kinokuniya, Tokyo, 1982).

%%%  Nonlinear integrable systems 1981
\no{2}{Date, E., Jimbo, M., Kashiwara, M., and Miwa, T.,}
in \book{Nonlinear Integrable Systems}, M. Jimbo and T. Miwa (eds.)
(World Scientific, Singapore, 1983).

%%%  dispersionless Lax equations and topological minimal models
\no{3}{Krichever, I.M.,}
The dispersionless Lax equations and topological
minimal models, Commun. Math. Phys. (to appear);
Topological minimal models and soliton equations, reported at
the 1st Sakharov conference, May 1991.

%%% topological minimal models
\no{4}{Dijkgraaf, R., Verlinde, H., and Verlinde, E.,}
Topological strings in $d<1$,
Princeton preprint PUPT-1204, ASSNS-HEP-90/71 (October, 1990).

%%% topological minimal models
\no{5}{Blok, B., and Varchenko, A.,}
Topological conformal field theories and flat coordinates,
Princeton preprint IASSNS-HEP-91/05 (January, 1990).

%%% H-spaces with rotational Killing vector
\no{6}{Boyer, C., and Finley, J.D.,}
\jr{J. Math. Phys.} \vl{23} (1982), 1126-1128.

%%% H-spaces with rotational Killing vector
\no{7}{Gegenberg, J.D., and Das, A.,}
\jr{Gen. Rel. Grav.} \vl{16} (1984), 817-829.

%%% Lax equation with Poisson bracket structure
\no{8}{Golenisheva-Kutuzova, M.I., and Reiman, A.G.,}
\jr{Zap. Nauch. Semin. LOMI} \vl{169} (1988), 44 (in Russian).

%%% Lie algebra with Cartan operators
\no{9}{Saveliev, M.V., and Vershik, A.M.,}
\jr{Commun. Math. Phys.} \vl{126} (1989), 367-378.

%%% 3-d Toda, Lax formalism and solution formula
\no{10}{Kashaev, R.M., Saveliev, M.V., Savelieva, S.A., and Vershik, A.M.,}
On nonlinear equations associated with Lie algebras of diffeomorphism
groups of two-dimensional manifolds,
Institute for High Energy Physics preprint 90-I (1990).

%%% structure of W_\infty algebra (application to 3-d Toda)
\no{11}{Bakas, I.,}
\jr{Phys. Lett.} \vl{228B} (1989), 57-63;
\jr{Commun. Math. Phys.} \vl{134} (1990), 487-508.

%%% large-N limit of sigma model (application to 3-d Toda)
\no{12}{Park, Q-Han,}
\jr{Phys. Lett.} \vl{236B} (1990), 429-432;
\jr{Phys. Lett.} \vl{238B} (1990), 287-290.

%%% 3-d Einstein-Weyl
\no{13}{Hitchin, N.J.,}
in \book{Twistor Geometry and Non-linear Systems},
H.D. Doebner and T. Weber (eds.), Lecture Notes in Mathematics  vol. 970
(Springer-Verlag 1982).

%%% 3-d Einstein-Weyl
\no{14}{Jones, P.E., and Tod, K.P.,}
 \jr{Class. Quantum Grav.} \vl{2} (1985), 565-577.

%%% 3-d Einstein-Weyl
\no{15}{Ward, R.S.,}
\jr{Class. Quantum Grav.} \vl{7} (1990). L95-L98.

%%% 3-d scalar flat K\"ahler geometry and minitwistor theory
\no{16}{LeBrun, C.,}
Explicit self-dual metrics on $CP_2$ \# $\dots$ \# $CP_2$,
\jr{J. Diff. Geometry} (to appear).

%%% SDiff(2) Toda hierarchy
\no{17}{Takasaki, K., and Takebe, T.,}
SDiff(2) Toda equation -- hierarchy, tau function, and symmetries,
\jr{Lett. Math. Phys.} (to appear).

%%% Orlov's Lax formalism
\no{18}{Grinevich, P.G., and Orlov, A.Yu.,}
in \book{Problems of modern quantum field theory},
A. Belavin et al. (eds.)
(Springer-Verlag, 1989).

%%% d=1 string, \hat{W}_\infty algebra, Orlov theory
\no{19}{Awada, M., and Sin, S.J.,}
Twisted $W_\infty$ symmetry of the KP hierarchy
and the string equation of $d=1$ matrix models,
Florida preprint UFITFT-HEP-90-33 (November, 1990)

%%% equations of KdV type
\no{20}{Segal, G., and Wilson, G.,}
\jr{Publ. IHES} \vl{61} (1985), 5-65.

%%% conservation law of KP
\no{21}{Cherednik, I.V.,}
Funct. Anal. Appl. 12 (1978), 195-203.

%%% conservation law of KP
\no{22}{Wilson, G.,}
Quart. J. Math. Oxford 32 (1981), 491-512.

%%% conservation law of KP
\no{23}{Flaschka, H.,}
Quart. J. Math. Oxford 34 (1983), 61-65.

%%% An infinite number of ..., JMP
\no{24}{Takasaki, K.,}
\jr{J. Math. Phys.} \vl{30} (1989), 1515-1521.

%%% flat coordinates
\no{25}{Ishiura, S., and Noumi, M.,}
\jr{Proc. Japan Acad.} \vl{58} (1982), 13, 62.

%%% nonlinear graviton construction
\no{26}{Penrose, R.,}
\jr{Gen. Rel. Grav.} \vl{7} (1976), 31-52.

%%% hyper-K\"ahler geometry
\no{27}{Hitchin, N.J., Kahlhede, A., Lindstr\"om, U., and Ro\v{c}ek, M.,}
\jr{Commun. Math. Phys.} \vl{108} (1987), 535-589.

%%% polygons and gravitons
\no{28}{Hitchin, N.J.,}
\jr{Math. Proc. Camb. Phil. Soc.} \vl{85} (1979), 465-476.

%%% nonlinear graviton with s.d. Killing vector
\no{29}{Tod, K.P., and Ward, R.S.,}
\jr{Proc. R. Soc. London} \vl{A363} (1979), 411-427.

%%% hydrodynamic Hamiltonian structure
\no{30}{Dubrovin, B.A., and Novikov, S.P.,}
\jr{Soviet Math. Dokl.} \vl{27} (1983), 665-669.

%%% hydrodynamic Hamiltonian structure
\no{31}{Tsarev, S.P.,}
\jr{Soviet Math. Dokl.} \vl{31} (1985), 488-491.

%%% averaging method
\no{32}{Krichever, I.M.,}
\jr{Funct. Anal. Appl.} \vl{22} (1989), 200-213.

%%% [33] Gelfand-Dikii resolvent functional for KdV
\no{33}{Gel'fand, I.M., and Dikii, L.A.,}
\jr{Russian Math. Surveys} \vl{30:5} (1975), 77-113.

%%% Virasoro constraints
\no{34}{Dijkgraaf, R., Verlinde, E., and Verlinde, H.,}
Loop equations and Virasoro constraints in non-perturbative
2d quantum gravity,
\jr{Nucl. Phys.} \vl{B348} (1991), 435-456.
%%Princeton preprint PUPT-1184, IASSNS-HEP-90/48 (1990).

%%% Virasoro constraints
\no{35}{Fukuma, M., Kawai, H., and Nakayama, R.,}
Continuum Schwinger-Dyson equations and universal structures in
two-dimensional quantum gravity,
\jr{Int. J. Mod. Phys.} \vl{A6} (1991), 1385-1406.
%%%Tokyo preprint UT-562 (May 1990).

%%% geometry of string equations
\no{36}{Moore, G.,}
\jr{Commun. Math. Phys.} \vl{133} (1990), 261-304.

%%% An infinite ... JMP
\no{37}{Boyer, C.P., and Plebanski, J.F.,}
\jr{J. Math. Phys.} \vl{26} (1985), 229-234.

%%% Symmetries of hyper-K\"ahler ... JMP
\no{38}{Takasaki, K.,}
\jr{J. Math. Phys.} \vl{31} (1990), 1877-1888.

%%% differential algebra, D-module and STL hierarchy
\no{39}{Takasaki, K.,}
\jr{Lett. Math. Phys.} \vl{19} (1990), 229-236.

%%% cocycles of SDiff(2)
\no{40}{Arakelyan, T.A., and Savvidy, G.K.,}
\jr{Phys. Lett.} \vl{214B} (1988), 350-356.

%%% cocycles of SDiff(2)
\no{41}{Bars, I., Pope, C.N., and Sezgin, E.,}
\jr{Phys. Lett.} \vl{210B} (1988), 85-91.

%%% cocycles of SDiff(2)
\no{42}{Floratos, F.G., and Iliopoulos, J.,}
\jr{Phys. Lett.} \vl{201B} (1988), 237-240.

%%% cocycles of SDiff(2)
\no{43}{Hoppe, J.,}
\jr{Phys. Lett.} \vl{215B} (1988), 706-710.

%%% dispersionless KP and Toda hierarchy, special solutions, etc
\no{44}{Kodama, Y.,}
A method for solving the dispersionless KP equation
and its exact solutions,
\jr{Phys. Lett.} \vl{129A} (1988), 223-226;
%%%
Kodama, Y., and Gibbons, J.,
A method for solving the dispersionless KP hierarchy
and its exact solutions, II,
\jr{Phys. Lett.} \vl{135A} (1989), 167-170;
%%%
Kodama, Y.,
Solutions of the dispersionless Toda equation,
\jr{Phys. Lett.} \vl{147A} (1990), 477-482.

\bye